%% file: draft_main.tex
\documentclass[fleqn,usenatbib,useAMS]{mnras}

\usepackage{newtxtext,newtxmath}

\usepackage[T1]{fontenc}

\DeclareRobustCommand{\VAN}[3]{#2}
\let\VANthebibliography\thebibliography
\def\thebibliography{\DeclareRobustCommand{\VAN}[3]{##3}\VANthebibliography}

\usepackage{graphicx}	
\usepackage{amsmath}	

\usepackage{epstopdf}

\title[]{Giant Molecular Cloud Formation at the Interface of Colliding Supershells in the Large Magellanic Cloud}

\author[K. Fujii et al.]{
Kosuke Fujii$^{1,2,3}$\thanks{E-mail: kosuke.fujii@nao.ac.jp},
Norikazu Mizuno$^{1,2}$, 
J. R. Dawson$^{3,4}$, 
Tsuyoshi Inoue$^{5}$, 
\newauthor
Kazufumi Torii$^{6}$, 
Toshikazu Onishi$^{7}$, 
Akiko Kawamura$^{2}$, 
Erik Muller$^{2}$, 
\newauthor
Tetsuhiro Minamidani$^{2,8}$, 
Kisetsu Tsuge$^{5}$
and Yasuo Fukui$^{5}$
\\
$^{1}$Department of Astronomy, School of Science, 
The University of Tokyo, 7-3-1 Hongo, Bunkyo-ku, Tokyo 133-0033, Japan\\
$^{2}$National Astronomical Observatory of Japan (NAOJ), National Institutes of Natural Sciences (NINS), 2-21-1, Osawa, Mitaka, Tokyo 181-8588, JAPAN\\
$^{3}$Australia Telescope National Facility, CSIRO Astronomy and Space Science, PO Box 76, Epping, NSW 1710, Australia\\
$^{4}$Department of Physics and Astronomy and MQ Research Centre in Astronomy, Astrophysics and Astrophotonics, Macquarie University, NSW 2109, Australia\\
$^{5}$Department of Astrophysics, Nagoya University, Furo-cho, Chikusa-ku, Nagoya 464-8602, Japan\\
$^{6}$Nobeyama Radio Observatory, 462-2 Nobeyama Minamimaki-mura, Minamisaku-gun, Nagano 384-1305\\
$^{7}$Department of Physical Science, Osaka Prefecture University, Gakuen 1-1, Sakai, Osaka 599-8531, Japan\\
$^{8}$Department of Astronomical Science, The Graduate University for Advanced Studies, SOKENDAI, 2-21-1, Osawa, Mitaka, Tokyo 181-8588, JAPAN
}

\date{Accepted 2021 April 16. Received 2021 April 13; in original form 2020 May 26.}

\pubyear{2021}

\begin{document}
\label{firstpage}
\pagerange{\pageref{firstpage}--\pageref{lastpage}}
\maketitle

\begin{abstract}
 	\input{draft_00abstract}

\end{abstract}
\begin{keywords}
ISM: clouds, ISM: bubbles, ISM: structure, Magellanic Clouds, galaxies: star formation, radio lines: ISM
\end{keywords}



\section{Introduction}
	\input{draft_01introduction}
\section{Observations}
	\input{draft_02obs}

\section{Results of the H$\,${\sc i} Observation}
	\input{draft_03results}

\section{Analysis\label{04} }
	\input{draft_04analysis}

\section{Discussion \label{05}}
	\input{draft_05discussion}
\section{Summary}
	\input{draft_06summary}




\bibliographystyle{mnras}
\bibliography{thesis}








\bsp	
\label{lastpage}
\end{document}

%% file: draft_00abstract.tex
We investigate the H$\,${\sc i} envelope of the young, massive GMCs in the star-forming regions N48 and N49, which are located within the high column density H$\,${\sc i} ridge between two kpc-scale supergiant shells, LMC 4 and LMC 5.
New long-baseline H$\,${\sc i} 21 cm line observations with the Australia Telescope Compact Array (ATCA) were combined with archival shorter baseline data and single dish data from the Parkes telescope, for a final synthesized beam size of 24.75$^{\prime \prime}$ by 20.48$^{\prime \prime}$, which corresponds to a spatial resolution of $\sim$ 6 pc in the LMC.
It is newly revealed that the H$\,${\sc i} gas is highly filamentary, and that the molecular clumps are distributed along filamentary H$\,${\sc i} features.
In total 39 filamentary features are identified and their typical width is $\sim$ 21 (8--49) [pc].
We propose a scenario in which the GMCs were formed via gravitational instabilities in atomic gas which was initially accumulated by the two shells and then further compressed by their collision. 
This suggests that GMC formation involves the filamentary nature of the atomic medium.

%% file: draft_01introduction.tex
Giant molecular clouds (GMCs) are the principal sites of stellar cluster formation; most of the stars in a galaxy are formed within them.
Understanding how GMCs are formed via the conversion of H$\,${\sc i} into H$_2$, and how they evolve to form
stars and star clusters, is therefore of critical importance in order to build a general understanding of both star formation and galaxy evolution.

Recent theoretical works on GMC formation have argued that GMCs can form from a diffuse ambient medium via multiple shock compressions (\citealt{Inoue_inutsuka_2012, Inutsuka_etal_2015}).
Simulations of supersonic converging flows have shown that the action of the thermal instability at the flow stagnation point can generate dense clouds by piling up the warn neutral medium (WNM; e.g. \citealt{Hennebelle_Perault_1999, Koyama_Inutsuka_2000, Koyama_Inutsuka_2002, Audit_Hennebelle_2005, Hennebelle_Audit_2007, Heitsch_etal_2005, Heitsch_etal_2006, Ntormousi_etal_2011, Inoue_inutsuka_2012}).
When high-resolution magneto-hydrodynamical simulations are performed, the magnetic pressure acts to oppose the contraction of the clouds \citep{Inoue_Inutsuka_2008, Inoue_Inutsuka_2009, Heitsch_etal_2009, Ntormousi_etal_2017}, so that the formation of dense clouds of $n > 10^3$ cm$^{-3}$ typically requires multiple episodes of supersonic compression (\citealt{Inutsuka_etal_2015}). 
In this context, it is important to study not only the 
GMCs themselves, but also their H{\sc i} envelopes, which may contain clues to their assembly from the atomic ISM. 

The kinematics of HI envelopes in particular can provide important information on any ongoing accretion.  
However, to-date there has been almost no work focusing on the kinematics of the atomic envelopes of GMCs with sufficient spatial resolution. 
Nevertheless, several studies have provided an important foundation.
\cite{Fukui_etal_2009} have performed a statistical analysis of the dynamical correlation between GMCs and their H$\,${\sc i} envelopes using the entire population of GMCs in the Large Magellanic Cloud (LMC), observed in CO at a resolution of 40 pc.
These authors find that the GMCs likely grow by accreting mass from their filamentary H$\,${\sc i} envelopes.
Further analysis, especially observations at higher spatial resolution, is now required since the GMCs identified at 40 pc resolution typically consist of many sub-component clouds with characteristic sizes of $\sim$ 10 pc (\citealt{Wong_etal_2011}).
On smaller, sub-pc scales, recent ALMA observations the LMC N159 region have revealed that LMC GMCs are also highly filamentary, similar to the filamentary clouds observed in the Milky Way (\citealt{Fukui_etal_2015c, Saigo_etal_2016}).
In order to understand the origin of such filamentary molecular clouds, it is important to increase the number of high-resolution case studies of the atomic environment of young GMCs.

The LMC star forming regions N48 and N49 (\citealt{Henize_1956}) are located between two optically identified Super Giant Shells (SGSs), LMC 4 and LMC 5 (\citealt{Meaburn_1980}), 
which are thought to have been formed by multiple generations of stellar feedback. 
Collisions between supershells (and super giant shells) have been proposed as a likely means of driving the multiple episodes of shock compression required to form GMCs from a magnetised medium \citep{Inutsuka_etal_2015}.
LMC 4 is the largest SGS in the LMC, with a size of $1.0 \times 1.8$ kpc in H$\alpha$, and LMC 5 is located northwest of LMC 4 with a diameter of $\sim$ 800 pc in H$\alpha$ (\citealt{Meaburn_1980}). 
The N48 and N49 regions contain two young, massive GMCs, with a total combined mass of $\sim$ 1.5$\times 10^{6}$ M$_{\odot}$ (\citealt{Mizuno_etal_2001, Fukui_etal_2008}). 
These GMCs are located within a high column density H$\,${\sc i} envelope that forms a giant ridge between the two SGSs, more than 500 pc long (hereafter the H$\,${\sc i} ridge).
The molecular clouds are believed to have formed within this H$\,${\sc i} ridge, which was presumably swept up by the two SGSs (\citealt{Yamaguchi_etal_2001a, Mizuno_etal_2001}). 
Recent ASTE $^{12}$CO($J$=3--2) observations by \citet{Fujii_etal_2014} have found that the GMCs have quite a clumpy distribution. These authors argue that the large-scale structure of the SGSs, especially the interaction between them, has led to the efficient formation of dense molecular clumps and stars.

The H$\,${\sc i} ridge of the N48/N49 region is therefore an ideal 
target to investigate the characteristic kinematics of the atomic envelope of a GMC formed via shock compression, as well as the link between these local kinematics and the large-scale dynamics of the expanding shells (which are mostly atomic gas). To this end, high resolution observations of the H$\,${\sc i} 21 cm line were carried out. 
In order to construct a plausible formation scenario for the GMCs in the H$\,${\sc i} ridge, the questions to be answered in this work are as follows: 
\begin{itemize}
\item What are the morphological characteristics of the GMCs and their H$\,${\sc i} envelope? Are they clumpy, filamentary, or clumpy filamentary? Do they have a more complicated distribution?
\item How were the GMCs formed in the SGS collision zone? Is collision-induced instability alone sufficient for their formation?
\end{itemize}

This work is the first case study at high spatial resolution around these GMCs. The outline of the paper is as follows. The details of the observations are described in \S 2. In \S 3, the main observational results are presented, including channel maps, and the detailed morphological characteristics of the ridge as seen in high resolution H$\,${\sc i}. 
In \S 4, three physical analyses of the H$\,${\sc i} data are introduced. The first is a correction for H{\sc i} opacity using archival dust opacity data; the second is the identification of the filamentary features seen in the high-resolution H$\,${\sc i} channel maps; the third is an analysis of the large scale gas dynamics of the ridge, as revealed by position velocity diagrams.
Finally, in \S 5, a possible GMC formation scenario for this region is proposed and discussed, and the work is summarized in \S 6.

%% file: draft_02obs.tex
The Australia Telescope Compact Array (ATCA) is an array of six 22m diameter antennas located at the Paul Wild Observatory near Narrabri in rural New South Wales, which is operated as part of the Australia Telescope National Facility (ATNF). 
At the time this study was carried out, it was the only interferometer capable of observing the 21cm line of atomic Hydrogen in the Southern Hemisphere.

\subsection{Observing Strategy \label{c2_hiobs}}
Observations were performed towards the N48 and N49 regions on January 27th  (1.5B configuration); May 1st (1.5D); and November 7th, 2014 (1.5A). The targeted region was mosaicked in 3 pointing centers separated by 16.8 arcminutes, sufficiently Nyquist sampling the 34$^{\prime}$ beam.
The observing time per day was 12 hours (January 27th, November 7th) and 10 hours (May 1st), which provides excellent $u$-$v$ coverage for each complementary array configuration.
Each pointing centre was observed for 20 seconds at a time, with the full set of three positions observed 20 times between each standard gain/phase calibrator scan.
The primary calibrator, PKS B1934-638 (assumed flux density 14.9 Jy at 1.419 GHz), was observed for bandpass and absolute flux-density calibration with 5 minutes integration at the start of each observing day.
The secondary calibrators, PKS B0515-674 (assumed flux density 1.18 Jy at 1.419 GHz) and PKS B0407-658 (assumed flux density 15.5 Jy at 1.419 GHz) were observed for gain and phase calibration. PKS B0515-674 was observed every 40 minutes with a 5-minute integration time (January 27th), and PKS B0407-658 was observed every 60 minutes with a 2-minute integration time (May 1st, November 7th).
The Compact Array Broadband Backend (CABB) was tuned to the CFB 1M-0.5k mode, which provides a bandwidth of 2 GHz split into 2048 1-MHz channels in each of two IFs, and a fine resolution (channel separation) of 0.5 kHz (2048 channels across 1 MHz) in up to 16 zoom bands in each IF. The observing band was a zoom band centered on 1.419 GHz -- the frequency of the H{\sc i} line at the source velocity (local standard of rest) of 285 km s$^{-1}$ -- with a total velocity coverage of 211.3 km s$^{-1}$.
The intrinsic velocity resolution is 0.1 km s$^{-1}$.

Shorter baselines were provided by publicly available archival observations of the entire LMC 5 region (which includes N48 and N49), made with the EW352 and 750A array configurations (project code C2648, Ann-Mao, S.A., Dawson, J.R. et al.). 
Including these shorter baseline data sets, the total number of unique baselines is 42, ranging from 31 m to 1469 m.

Calibration and imaging were performed using standard routines from the ATNF {\it MIRIAD} software package (\citealt{Sault_Killeen_2009}). Calibration, zeroth-order polynomial baseline subtraction, and Doppler correction were carried out in the $u$--$v$ domain. The individual pointings were linearly combined and imaged using the {\it MIRIAD} task {\it INVERT}; a standard grid and fast Fourier transform technique. A Briggs visibility weighting robustness parameter of 0.5 was adopted. 
Deconvolution was performed using the maximum entropy-based deconvolution algorithm {\it MOSMEM} (\citealt{Sault_etal_1996}), and the images were restored using the task {\it RESTOR}. 
The pixel size of the final image is 10$^{\prime \prime}$, and the velocity channel width is 0.4 km s$^{-1}$. The final synthesized beam size is 24.75 by 20.48 arcsec with a position angle of $-$35 degrees.

Zero spacing data was provided by Parkes 64-m Dish archival data from the Galactic All-Sky Survey (GASS; \citealt{McClureGriffiths_etal_2009}, \citealt{Kalberla_etal_2010}). 
GASS is a 21-cm line survey covering the southern sky for all declinations $\delta \lesssim  1^{\circ}$, including the entire LMC, with a velocity range of $\left| v_{\rm lsr} \right| \lesssim 468$ km s$^{-1}$. The intrinsic beam size of the data is 14.4$^{\prime}$ (FWHM) and the effective velocity resolution is 1.0 km s$^{-1}$, with a channel width of $0.8$ km s$^{-1}$.
The GASS data were regridded to match the ATCA data, and the two were then linearly combined in the Fourier domain using the {\it MIRIAD} task {\it IMMERGE}. In order to perform this combination, 
it was necessary to ensure that the flux calibration of the two datasets was consistent --
typically achieved by assuming that the interferometer data has the better absolute calibration, and applying an appropriate scaling factor to the single-dish data in order to match the two.
This scaling factor was estimated by examining data in the region of the UV plane that is well measured by both the mosaicked ATCA and Parkes observations, which was assumed to be an annulus covering the range 20--50 m. 
The comparison was performed over a rectangular area of $\sim$ 34$^{\prime}$ $\times$ 52$^{\prime}$ ($\sim500 \times 800$ pc) encompassing the HPBW field of the three high-resolution mosaic pointings (bottom left position [5h 29m 30.83s:$-$66$^{\circ}$ 35$^{\prime}$ 47.7$^{\prime \prime}$], top right position [5h 22m 9.16s:$-$65$^{\circ}$ 44$^{\prime}$ 44.7$^{\prime \prime}$]).
A reasonable scaling factor of 1.125 was found, and applied to the GASS data prior to merging. The effective velocity resolution of the final merged cube is limited by the GASS data, and is 1.0 km s$^{-1}$.

\subsection{Flux Consistency}

For the LMC, archival ATCA+Parkes H$\,${\sc i} survey data exists (\citealt{Kim_etal_2003}), consisting of the ATCA observations of \citet{Kim_etal_1998}, combined with the Parkes data of \citet{Staveley-Smith_etal_2003} via an image feathering (linear merging) approach. This survey cube covers the entire LMC at a spatial resolution of 60$^{\prime \prime}$ and a pixel size of 20$^{\prime \prime}$, and has a 1$\sigma$ noise level of 2.4 K in a 1.65 km s$^{1}$ velocity channel. While the telescopes used are the same than in the present work, the array configurations (and hence the $u$--$v$ coverage of the datasets) are different, and confirming flux consistency between the two is a useful check of our data.

Before comparison, the new high-resolution data was smoothed to a beam size of $60^{\prime \prime}$, and the archival data was regridded to match the new data's (10$^{\prime \prime}$) grid. The total velocity-integrated flux density of the new data, within the rectangular area defined in \S\ref{c2_hiobs} above, is $3.5 \times 10^5$ Jy/Beam km s$^{-1}$, compared to $3.9 \times 10^5$ Jy/Beam km s$^{-1}$ for the archival data -- a difference of greater than 10\%.
This is likely due to the presence of negative side-lobes in the new data. These arise from the large dynamic range of the H$\,${\sc i} emission in the small region covered by our three mosaic pointings, which encompasses both the bright ridge and the evacuated cavity of the supergiant shells. 
If these negative bowls are excluded, the total integrated flux density is consistent between the two data sets.

%% file: draft_03results.tex

\subsection{High-resolution H$\,${\sc i} Map \label{c3_HImap}}
\begin{figure*}
 \begin{center}
  \includegraphics[width=2\columnwidth]{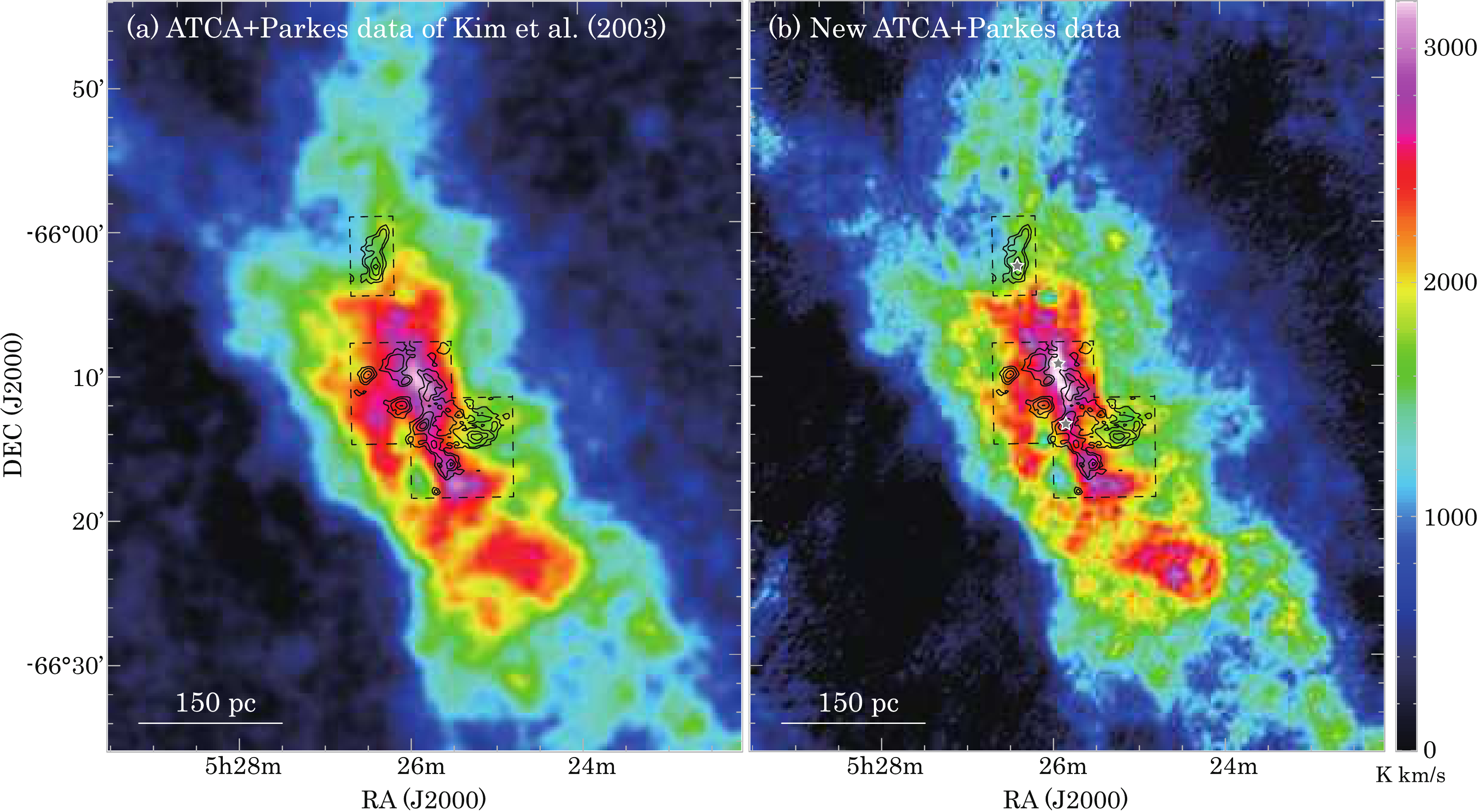}  
 \end{center}
\caption{
(a) Color map of the H$\,${\sc i} integrated intensity of \citet[][beam FWHM: 60$^{\prime \prime}$]{Kim_etal_2003}. Black contours are ASTE $^{12}$CO($J$=3--2) integrated intensity (\citealt{Fujii_etal_2014}), and black dotted boxes show the coverage of the ASTE observations.
The lowest contour is 4.0 K km s$^{-1}$ and thereafter run in steps of 8.0 K km s$^{-1}$.
(b) Color map of the H$\,${\sc i} integrated intensity of the new ATCA$+$Parkes data (beam FWHM: $24.7^{\prime \prime} \times 20.5^{\prime \prime}$). Contours and boxes are the same as (a). Grey stars, in the right panel, mark the positions of the spectra shown in Fig. \ref{fig_C3.HI.spec}.
\label{fig_C3.HI.mom0}
}  
\end{figure*}

Figure \ref{fig_C3.HI.mom0} presents integrated intensity maps of the archival ATCA$+$Parkes H$\,${\sc i} data \citep{Kim_etal_2003}, and our new high-resolution H$\,${\sc i} data. The angular resolutions of 60$^{\prime \prime}$, and 24.7$^{\prime \prime} \times 20.5^{\prime \prime}$, respectively, correspond to spatial resolutions of 15 pc and $6 \times 5$ pc at the distance of the LMC.
The resolution is quite high for atomic gas around star forming regions in external galaxies, and the $\sim$5 pc resolution of the new data is slightly higher than the FWHM of the ASTE $^{12}$CO($J$=3--2) data ($\sim$ 27$^{\prime \prime}$, $\sim$ 8 pc) previously presented in \citet{Fujii_etal_2014}. On large scales, the H$\,${\sc i} intensity distributions of the two images are very similar, confirming that the new observations successfully reproduce previous results.
On small scales, the new data resolve significantly finer structure.
Although the central part of the ridge still contains some obviously unresolved structure, the more diffuse material surrounding the ridge is clearly far better resolved than in the older data.

\begin{figure}
 \begin{center}
  \includegraphics[width=\columnwidth]{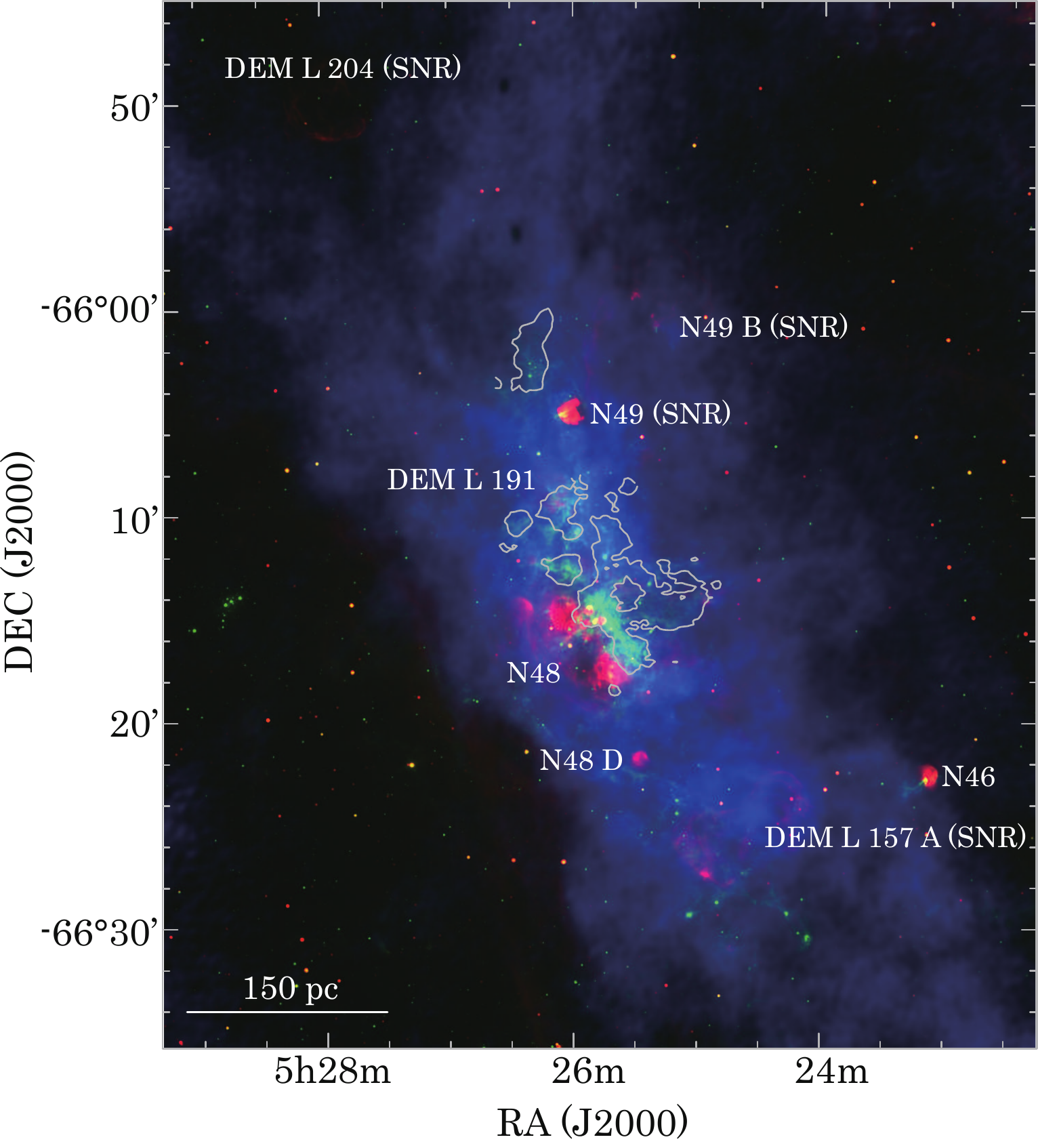}  
 \end{center}
\caption{
3 color image of the H$\,${\sc i} ridge.
Red is H$\alpha$ (\citealt{MCELS_1999})
, green is Spitzer 8.0 $\mu$m data (\citealt{Meixner_etal_2006})
, and blue is the H$\,${\sc i} data of the present work.
Gray contours are the lowest contour of $^{12}$CO($J$=3--2) shown in Figure \ref{fig_C3.HI.mom0}.
\label{fig_C3.HI.3col}}  

\end{figure}

Figure \ref{fig_C3.HI.3col} shows a comparison between the H$\,${\sc i} and tracers of star formation activity. 
H$\alpha$ data is from the Magellanic Cloud Emission-Line Survey (MCELS; \citealt{MCELS_1999}), and 8 $\mu$m data is taken from the Spitzer Legacy Program ``SAGE'' \citep{Meixner_etal_2006}. 
The names of H$\,${\sc ii} regions and supernova remnants identified by \cite{Henize_1956} and \cite{Davies_Meaburn_1976} are indicated in the figure. 
The improved resolution is quite clear in the region of the H$\,${\sc i} holes containing the N49 SNR ($\sim$ 5h 26m:$-$66$^{\circ}$ 05$^{\prime}$) and the H$\,${\sc ii} region N48 ($\sim$ 5h 26m:$-$66$^{\circ}$ 25$^{\prime}$). The other H$\alpha$ emission regions do not show any clear correspondance with features in the H$\,${\sc i} integrated intensity image.

Two clear absorption features associated with compact background continuum sources are newly detected in the north part of the image ([5h 26m 25.9s; $-$65$^{\circ}$ 56$^{\prime}$ 19.0$^{\prime \prime}$] and [5h 26m 32.9s; $-$65$^{\circ}$ 49$^{\prime}$ 7.9$^{\prime \prime}$]).
These objects are not seen in the H$\alpha$ map, but they are identified as 1.4 GHz radio continuum sources N49 C and N49 D with a total flux of $\sim$ 0.2 Jy (\cite{Filipovic_etal_1998}).
H$\,${\sc i} absorption toward these sources has not been reported except for \cite{MarxZimmer_etal_2000}, who mention that one H$\,${\sc i} absorption feature is detected around this area.
The angular sizes of the two absorption features are slightly larger than the beam size of the new data ($\sim$ $40 ^{\prime \prime} \times 60 ^{\prime \prime}$ and $\sim$ $40 ^{\prime \prime} \times 50 ^{\prime \prime}$, respectively), suggesting an intrinsic angular extent of $\sim$ $30 ^{\prime \prime}$ to $50 ^{\prime \prime}$, once the $\sim$ $25 ^{\prime \prime}$ beam size is accounted for.

\subsection{High-resolution H$\,${\sc i} Spectra and Channel Maps}
\begin{figure*}
 \begin{center}
  \includegraphics[width=2.1\columnwidth]{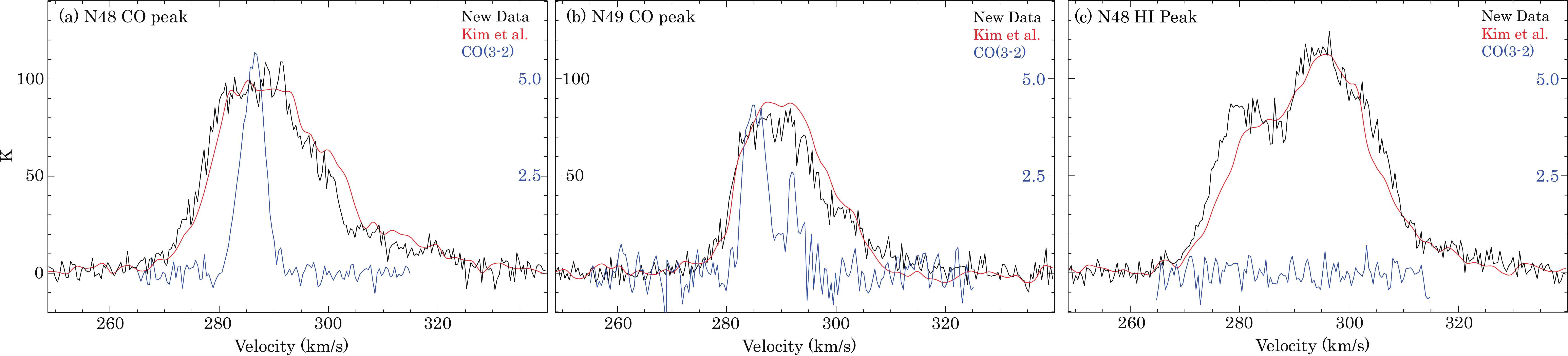}  
 \end{center}
\caption{
Examples of spectra from our new H$\,${\sc i} data (black), together those from the archival H$\,${\sc i} data (\citealt{Kim_etal_2003}; red), and ASTE $^{12}$CO($J$=3--2) data (blue). See the grey star marks shown in Figure \ref{fig_C3.HI.mom0} for the spatial positions of the spectra.
(a) Spectra at the strongest $^{12}$CO($J$=3--2) peak position in the N48 region (5h 25m 47.6s, -66$^{\circ}$ 13$^{\prime}$ 55.3$^{\prime \prime}$).
(b) Spectra at the $^{12}$CO($J$=3--2) peak position in the N49 region (5h 26m 18.8s, -66$^{\circ}$ 02$^{\prime}$ 51.5$^{\prime \prime}$).
(c) Spectra at the most luminous H$\,${\sc i} position in the ridge (in the N48 region, 5h 25m 51.9s, -66$^{\circ}$ 09$^{\prime}$ 34.8$^{\prime \prime}$).
\label{fig_C3.HI.spec}}   
\end{figure*}

Example spectra at the CO emission peaks of N48 and N49, and at the position of peak H$\,${\sc i} integrated intensity, are shown in Figure \ref{fig_C3.HI.spec}. 
It can be seen that in the vicinity of the CO peaks, the H$\,${\sc i} spectra from \citet{Kim_etal_2003} are similar in shape to our new data, with peak temperatures that only vary by a small amount in our improved resolution cubes (spatial and spectral).
This indicates that the atomic gas is relatively smoothly distributed in these regions. 
On the other hand, at the peak position of the H$\,${\sc i}, the new data shows a clearer double component -- i.e.
we are better resolving small-scale structure in the atomic gas at this position. 
The CO spectra are roughly centered around the peak velocities of the H$\,${\sc i} spectra, but their spectral shapes are quite different, 
presumably reflecting a complex and more spatially-extended distribution of H$\,${\sc i} around the CO clouds.

Figures \ref{fig_C3.HI.chan.0102} and \ref{fig_C3.HI.chan.0304} show velocity channel maps of the new data, 
binned to a channel width of 5.0 km s$^{-1}$.
The full velocity extent of the H$\,${\sc i} emission is 260 km s$^{-1}$ to 330 km s$^{-1}$, with
bright H$\,${\sc i} ($>$ 50 K; roughly the half of the peak intensity) seen between the 276 km s$^{-1}$ and 310 km s$^{-1}$ channels.
Each channel map shows complicated structures in the atomic gas, with a spatial distribution that varies significantly from panel to panel, 
indicating that the H$\,${\sc i} ridge contains many sub-structures. Most of these sub-structures appear quite filamentary, suggesting that the ridge might consist of a collection of filamentary structures. 
A detailed analysis of these filamentary features is carried out in \S \ref{c4_HIfilament}.

CO emission is detected between the 276 km s$^{-1}$ and 305 km s$^{-1}$ channels -- the range in which the H$\,${\sc i} is most luminous. 
In the 286 km s$^{-1}$ and 291 km s$^{-1}$ channels, which contain the majority of the CO emission, molecular clouds are clearly distributed along bright filamentary H$\,${\sc i} features.
This may be interpreted as evidence that the majority of the molecular gas has been formed in such luminous atomic filaments;
\textit{this is one of the most important suggestions of these new high-resolution H$\,${\sc i} observations.}
A potential exception is the N49 clouds (the Northern part of the region observed with ASTE), which -- while arguably lying along an atomic filament -- are not associated with bright H$\,${\sc i}. The weaker H$\,${\sc i} emission may possibly reflect a smaller quantity of diffuse atomic gas, a smaller H$\,${\sc i} scale height, or a higher molecular fraction. 
Conversely, there are several regions where H$\,${\sc i} is bright but molecular gas is not detected. 
This is not particularly surprising, since H$\,${\sc i} brightness temperature is not a good proxy for actual density;
in such regions, the number density of the atomic medium may not be high enough to form molecular clouds.

In the velocity range from 340 km s$^{-1}$ to 370 km s$^{-1}$, high-velocity H$\,${\sc i} components are detected from the northern part of the N48 region to the western part of the N49 region, offset by roughly 60 km s$^{-1}$ from the main emission components. The western part of N49 contains the supernova remnants N49 and N49B, suggesting that these high-velocity components may have been accelerated by the SNRs.

\begin{figure*}
 \begin{center}
  \includegraphics[width=1.8\columnwidth]{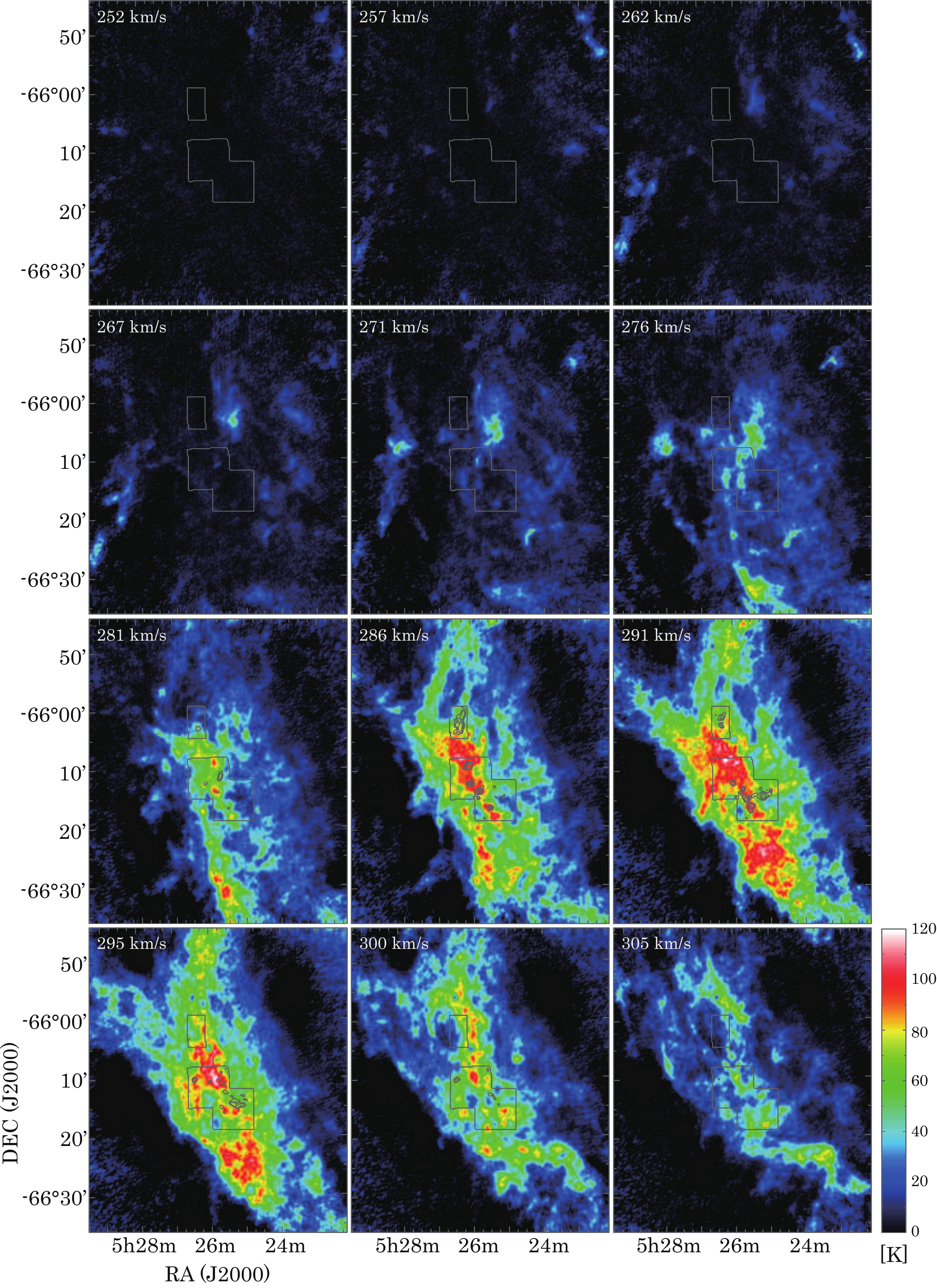}
 \end{center}
\caption{
Velocity channel maps of the H$\,${\sc i} brightness temperature of the new data (color) and $^{12}$CO($J$=3--2) brightness temperature (black contours) in the H$\,${\sc i} ridge (coverage is the same as Figure \ref{fig_C3.HI.mom0}), averaged over velocity intervals of 5.0 km s$^{-1}$. 
The central velocity is shown in the upper left of each panel. Black boxes indicated in all channels show the area observed with ASTE.
\label{fig_C3.HI.chan.0102}
}
\end{figure*}

\begin{figure*}
 \begin{center}
  \includegraphics[width=1.8\columnwidth]{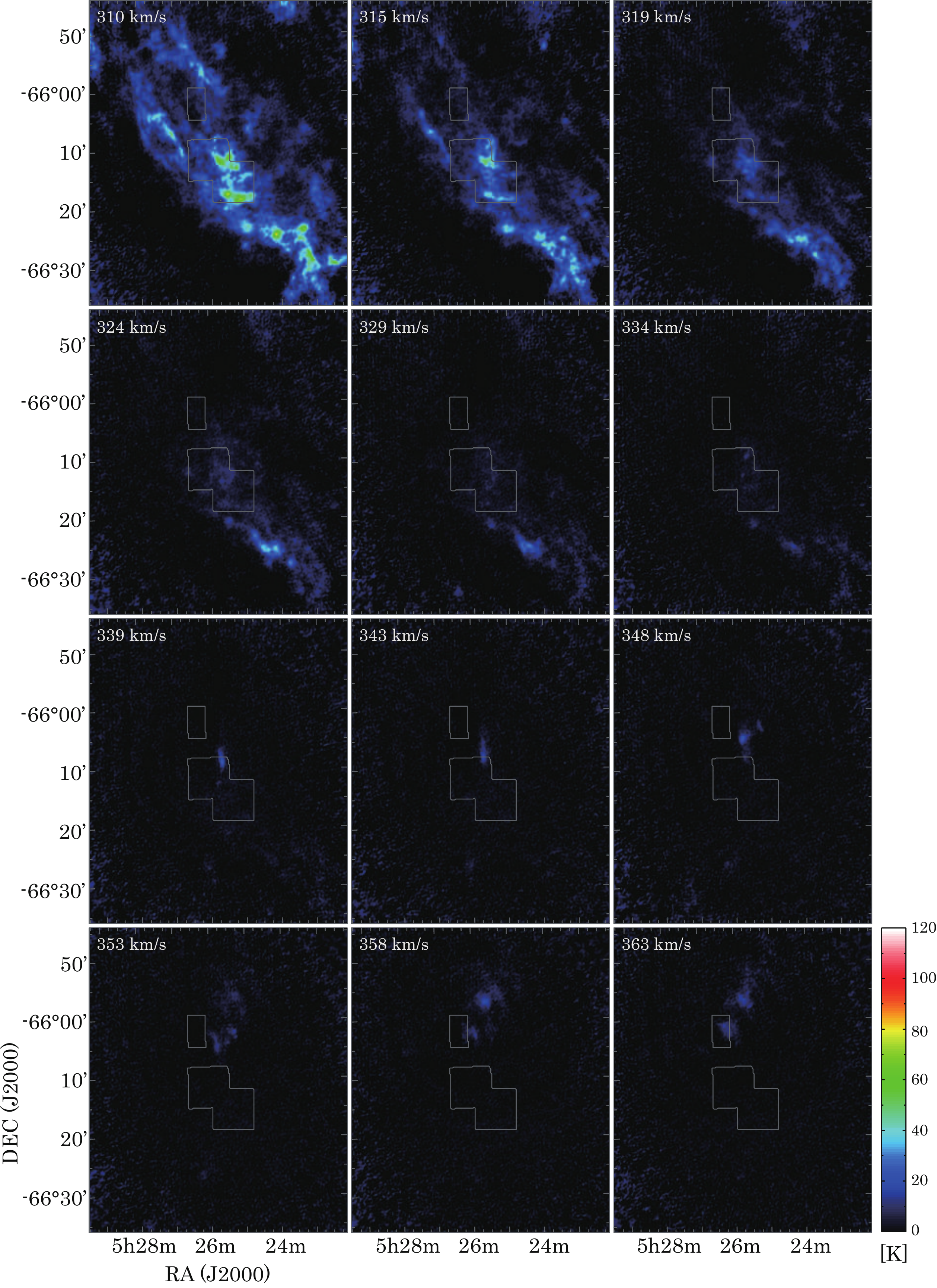}
 \end{center}
\caption{ 
(Continued channel maps)
\label{fig_C3.HI.chan.0304}}
\end{figure*}

%% file: draft_04analysis.tex
\subsection{Optically Thick H$\,${\sc i} \label{s4_opacity}}

We estimate opacity-corrected column densities and the total H$\,${\sc i} mass using the method devised by \cite{Fukui_etal_2014b} and \cite{Fukui_etal_2015a}. These authors use Planck dust opacity to derive an opacity-corrected H$\,${\sc i} column density (assuming that the gas and dust are well mixed with uniform properties), and treat H$\,${\sc i} as a single component to derive harmonic mean values for spin temperature and optical depth along the line of sight. Detailed descriptions of this methodology are provided in the papers cited above. As demonstrated by e.g. Lee et al. (2015), \cite{Murray_etal_2018} and \cite{Nguyen_etal_2018}, there is evidence that this method overestimates the total H$\,${\sc i} column density for Milky Way sight lines. Nevertheless, it is instructive to apply it here -- both as a simple measure of possible missing mass, and for the sake of comparing our results with those found in the Galaxy.

Archival datasets of dust optical depth at 353 GHz ($\tau _{353}$), dust temperature ($T _{\rm d}$), H$\,${\sc i}, and H$\alpha$ flux are used for this analysis. 
$\tau _{353}$ and $T _{\rm d}$ images were downloaded from the Planck Legacy Archive (PLA). These quantities were originally derived by fitting 353, 545, and 857 GHz data from the first 15 months of Planck satellite observations, together with 100 micron data from the IRAS satellite. The angular resolution of both datasets is 5 arcminutes, with a pixel size of $2 ^\prime \times 2 ^\prime$. For details, see the PLA explanatory supplement (\citealt{Planck_Collaboration_2014}).
Note that a Galactic foreground component has been subtracted from the Planck data in the same way as described in Section 3.7 of \cite{Tsuge_etal_2019}.
The H$\,${\sc i} data is obtained from the ATCA$+$Parkes survey of \citet{Kim_etal_2003}, smoothed with a Gaussian kernel in the image domain to a spatial resolution of 5 arcminutes.  
H$\alpha$ data is taken from the Magellanic Cloud Emission-Line Survey (MCELS; \citealt{MCELS_1999}), also smoothed to 5-arcminute resolution. The pixel size of all datasets is matched to the Planck data (2 arcminutes).
Since the LMC is almost face-on, pixel-to-pixel comparisons of 2D images can be made without significant line-of-sight contamination.

We also note that while higher resolution dust datasets do exist (e.g. \textit{Herschel} observations), the combination of Planck and IRAS data presents a significant advantage when converting to ISM mass, because the modified Planck function used to derive the dust temperature and optical depth is fit at four wavelengths that fully sample the peak of the function. 
The 5-arcmin resolution of the Planck/IRAS data is sufficient to resolve the dust distribution of the present GMC and H{\sc i} ridge.

\begin{figure*}
 \begin{center}
  \includegraphics[width=2.1\columnwidth]{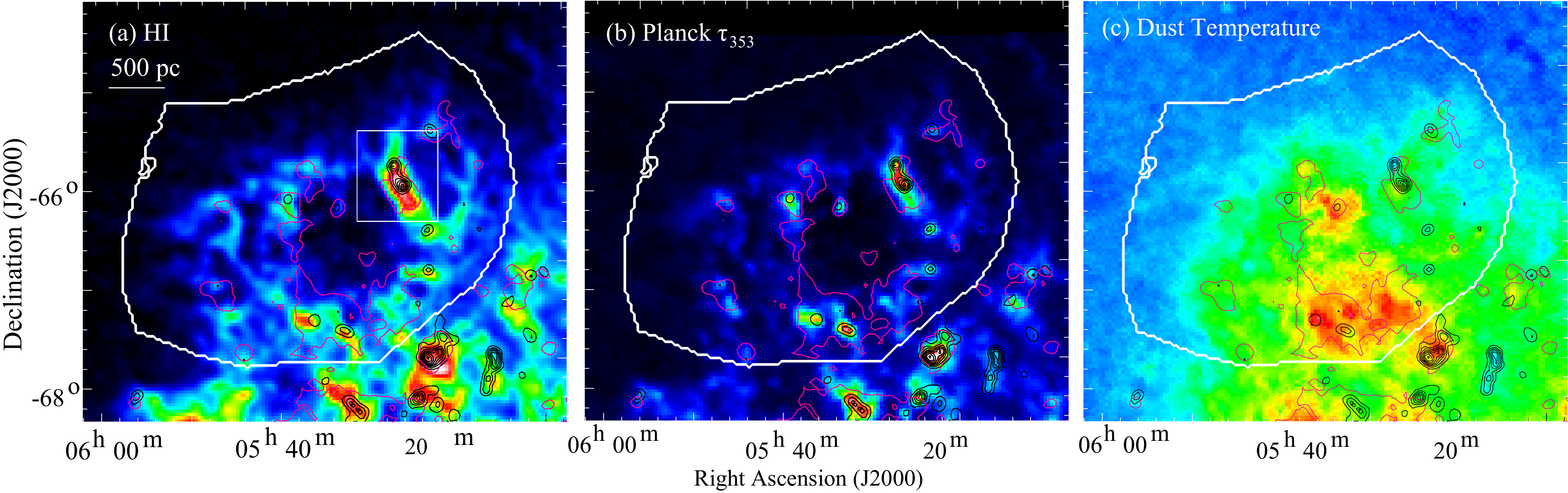}  
 \end{center}
\caption{
The three datasets used in the opacity-correction analysis.
(a) Color map of H$\,${\sc i} integrated intensity (\citealt{Kim_etal_2003}), smoothed to a resolution of 5$^{\prime}$. The positions of the central cavities of LMC 4 and LMC 5 are labelled.
(b) Color map of $\tau _{353}$ (\citealt{Planck_Collaboration_2014}). 
(c) Color map of dust temperature (\citealt{Planck_Collaboration_2014}).
In all panels, black contours are the NANTEN $^{12}$CO (\citealt{Fukui_etal_2008}, also smoothed to 5$^{\prime}$. The contours are 0.5 K km s$^{-1}$ $+$ 1.0 K km s$^{-1}$.). The solid white line encloses the region used in this analysis, and solid red lines enclose masked regions of significant H$\alpha$ flux ($> 3 \times 10^7 / 4 \pi$ [photons cm$^{-2}$ s$^{-1}$ Sr$^{-1}$]).
\label{fig_C4.HI.mask}}  

\end{figure*}

Figure \ref{fig_C4.HI.mask} shows the three datasets used in this analysis, and indicates the region over which it was carried out, roughly encompassing the entirety of LMC 4 and LMC 5.
We note that $\tau _{353}$ shows excellent spatial agreement with the CO data, arguably tighter than that between CO and H$\,${\sc i}. 
From the the dust temperature map it can be seen that $T _{\rm d}$ is high ($>$25 K) in the central and the southern parts of LMC 4, and moderate in other regions ($\sim$ 22 K on average).
In the area where heating by SNRs or the UV radiation from massive stars is significant, the relationship between the gas and dust properties may vary. 
In order to avoid including such regions in our analysis, we mask all areas where the H$\alpha$ flux is significant (indicated by closed red lines in Figure \ref{fig_C4.HI.mask}).
The masking threshold is set to $3 \times 10^7 / 4 \pi$ [photons cm$^{-2}$ s$^{-1}$ Sr$^{-1}$], which picks up all major H$\,${\sc ii} regions and SNRs in this area.

\begin{figure}
 \begin{center}
  \includegraphics[width=\columnwidth]{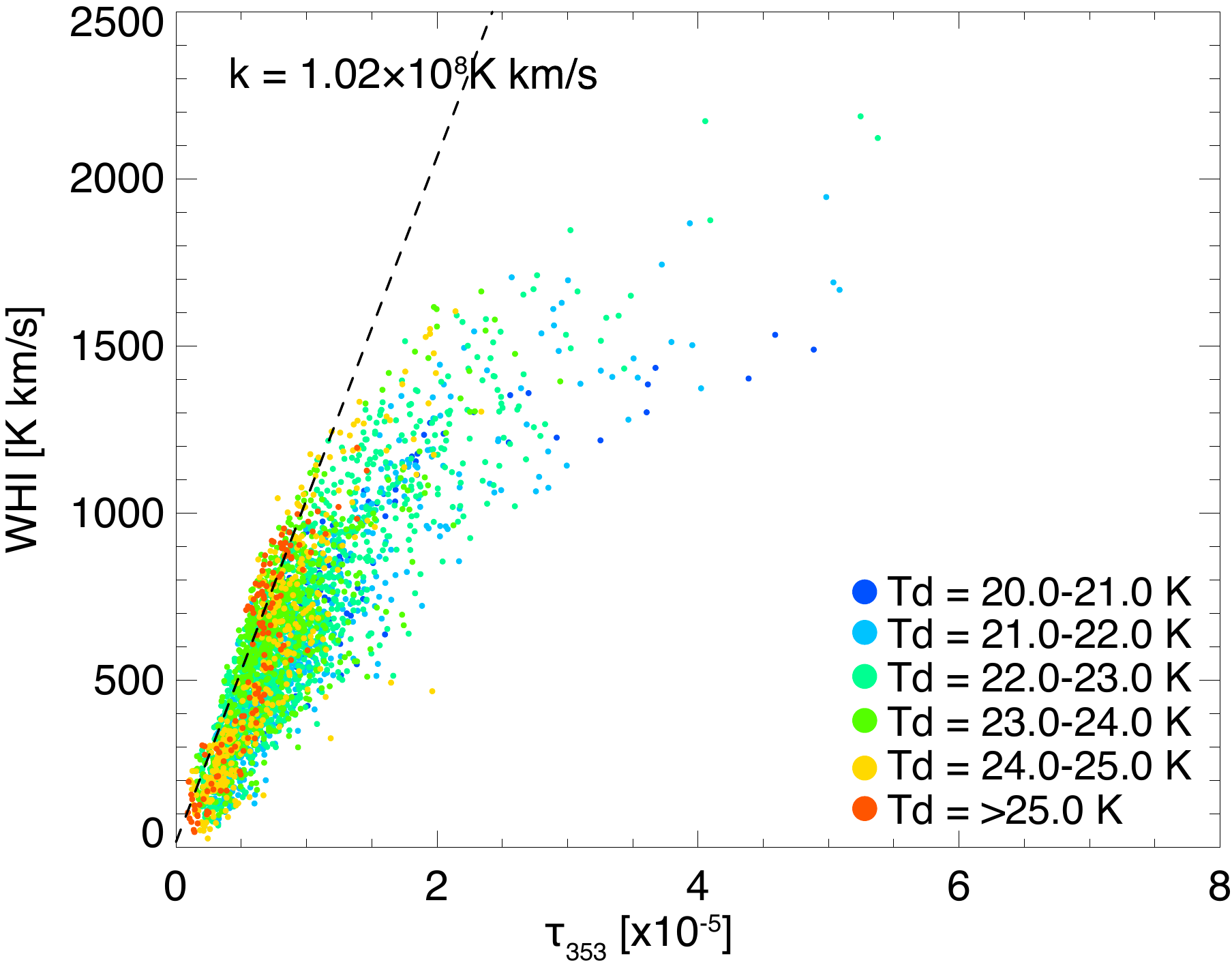}  
 \end{center}
\caption{
A scatter plot of $\tau _{353}$ and $W _{\rm HI}$ for the region defined in Figure \ref{fig_C4.HI.mask}. Plotted points are colored by their dust temperature, $T _{\rm d}$, in bins of 1 K, incremented from 20 K.
The dashed line is the line of best fit to the $T _{\rm d} > 25$ K datapoints, which has a slope, $k$, as indicated on the plot.
\label{fig_C4.HI.scaplot}}  

\end{figure}

A scatter plot of $\tau _{353}$ vs H$\,${\sc i} integrated intensity ($W _{\rm HI}$) is shown in Figure \ref{fig_C4.HI.scaplot}, with the data points color-coded by dust temperature. 
As in the Galactic case (\citealt{Fukui_etal_2014b, Fukui_etal_2015a}), the gas associated with the warmest dust (the T $>$ 25K bins) shows the clearest linear relationship between $\tau _{353}$ and $W _{\rm HI}$.
Following the papers cited above, we perform a least-squares linear fit to the T $>$ 25 K datapoints.
As shown in Figure \ref{fig_C4.HI.scaplot}, two components can be seen in the T $>$ 25 K plots, one with $W _{\rm HI} \lesssim $ 500 [K km s$^{-1}$], and the other with $W _{\rm HI} \gtrsim$ 500 [K km s$^{-1}$].
In this paper, the fit is performed on all data with $W _{\rm HI} > 500$ [K km s$^{-1}$], which gives the steepest slope i.e., the upper limit of the opacity correction analysis. 
The fit has a slope of $k = 1.02 \times 10^7$ [K km s$^{-1}$] and an intercept of $15.7$ [K km s$^{-1}$], with a correlation coefficient of 0.67. Note that a slope of entire T $>$ 25 K data is $k = 9.01 \times 10^6$, so the fit may include $\sim 10 \%$ error.

In the optically thin limit, H$\,${\sc i} column density is given exactly by 
\begin{equation}
N _{\rm HI, thin} = 1.823 \times 10 ^{18} \cdot W _{\rm HI}.
\label{eq_HIcolumn_whi}
\end{equation}
For high dust temperatures we may reasonably assume that we are in the optically thin regime, allowing us to obtain the appropriate scaling factor needed to derive $N _{\rm H}$ from $\tau _{353}$. 
Note that $\tau _{353}$ not only traces the true column density of H$\,${\sc i}, but is also tracing the H$_2$ in the molecular clouds (this is indeed expected under the assumption of a uniform gas-to-dust ratio).
Provided that dust properties are uniform, this relation will hold for all parameter space, allowing us to define an opacity-corrected column density $N _{\rm H, cor}$ as
\begin{equation}
N _{\rm H, cor} = (1.859 \times 10^{26}) \cdot \tau _{353},
\label{eq_HIcolumn_tau353}
\end{equation}
\noindent where $1.859 \times 10^{26}$ is simply the product of $1.823 \times 10^{18}$ and $k = 1.02 \times 10^7$ [K km s$^{-1}$].

In regions with no molecular gas, the spin temperature, $T _{\rm S}$, and optical depth, $\tau _{\rm HI}$, may then be calculated from the following coupled equations,
\begin{equation}
W _{\rm HI} \; [{\rm K \; km \; s}^{-1}] = (T _{\rm S} \; [{\rm K}] - T _{\rm bg} \; [{\rm K}]) \cdot \Delta V _{\rm HI} \; [{\rm km \; s}^{-1}] \cdot (1- \exp (- \tau _{\rm HI})) \label{eq_WHI_Ts}
\end{equation}
\begin{equation}
\tau _{\rm HI} = \frac{N _{\rm HI} \; [{\rm cm}^{-2}]}{1.823 \times 10 ^{18}} \cdot \frac{1}{T_{\rm S}} \cdot \frac{1}{\Delta V _{\rm HI} \; [{\rm km \; s}^{-1}]} \label{eq_tauHI}
\end{equation}
Here the H$\,${\sc i} linewidth, $\Delta V _{\rm HI}$, is computed as $W _{\rm HI}$/(peak H$\,${\sc i} brightness temperature), which assumes a single Gaussian emission component, 
and $T _{\rm bg}$ is the background continuum radiation temperature including the 2.7 K cosmic background radiation. 
And here $N _{\rm HI}$ is opacity-corrected column density of H$\,${\sc i}.
The resulting $T _{\rm S}$ and $\tau _{\rm HI}$ are in effect harmonic means, where it should also be noted that $\tau _{\rm HI}$ is an average over the H$\,${\sc i} velocity width $\Delta V _{\rm HI}$.
Equations (\ref{eq_WHI_Ts}) and (\ref{eq_tauHI}) correspond to two independent lines in the $T _{\rm S}$--$\tau _{\rm HI}$ plane.
We are therefore able to estimate $T _{\rm S}$ and $\tau _{\rm HI}$ from the crossing point of the two (see e.g. Figure 6 of \citealt{Fukui_etal_2014b}).
Note that in the optically thin limit, the two equations are essentially identical, and an infinite number of solutions are obtained. In this case, we may only obtain a lower limit for $T _{\rm S}$ and an upper limit for $\tau _{\rm HI}$.
We also note that the present method of averaging the optical depth over the linewidth has been examined by \cite{Fukui_etal_2018}. 
They have demonstrated that a line-averaged H$\,${\sc i} opacity and harmonic mean spin temperature well describe the behavior of H$\,${\sc i} gas based on synthetic observations of realistic, clumpy H$\,${\sc i} (\citealt{Inoue_inutsuka_2012}).
Roughly speaking, H$\,${\sc i} opacity is inversely proportional to $\Delta V$ for fixed $N _{\rm HI}$, because the CNM clumps will become more widely scattered in velocity for larger $\Delta V$.
Figure 13 of \cite{Fukui_etal_2018} also shows that the high opacity gas in real observations, i.e. CNM clumps, has a small area filling factor of less than $\sim 30 \%$, which matches the simulations.

\begin{figure*}
 \begin{center}
  \includegraphics[width=2\columnwidth]{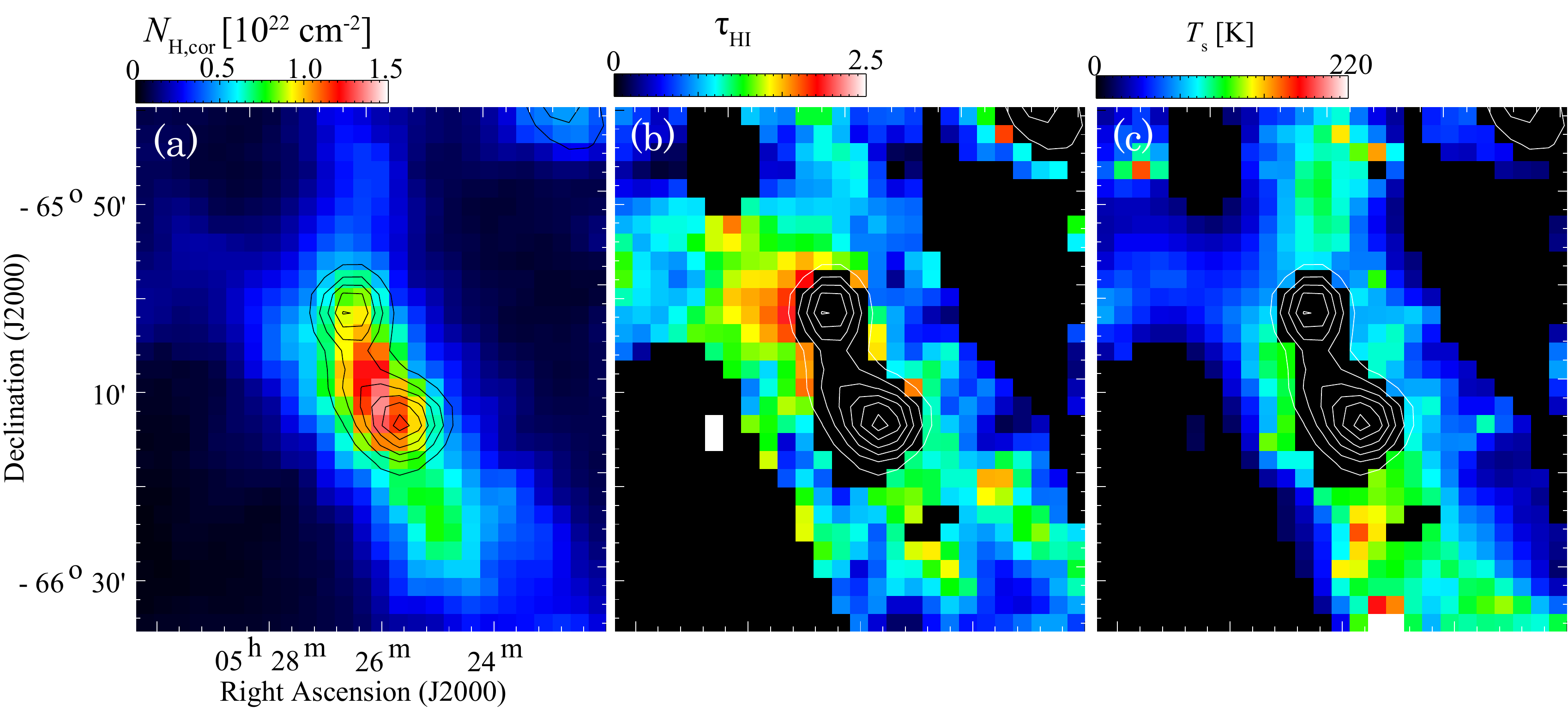}  
 \end{center}
\caption{
Three resulting maps of the opacity correction analysis.
(a) Color map of the modified total proton column density, $N _{\rm H, cor}$, as computed from the dust optical depth, $\tau _{353}$, using equation (\ref{eq_HIcolumn_tau353}) with $k = 9.74 \times 10^7$ [K km s$^{-1}$]. Contours are NANTEN $^{12}$CO smoothed to the 5$^{\prime}$ (starts from 0.5 K km s$^{-1}$ within steps of 1.0 K km s$^{-1}$). The area where the H$\,${\sc i} is too faint is masked.
(b) Color map of the H$\,${\sc i} optical depth $\tau _{\rm HI}$ derived from Equations (\ref{eq_WHI_Ts}) and (\ref{eq_tauHI}). Pixels of $^{12}$CO integrated intensity $> 0.5$ K km s$^{-1}$ and H$\alpha$ flux $> 6 \times 10^7 / 4 \pi$ [photons cm$^{-2}$ s$^{-1}$ Sr$^{-1}$] are masked.
(c) Color map of the H$\,${\sc i} spin temperature $T _{\rm S}$ derived from Equations (\ref{eq_WHI_Ts}) and (\ref{eq_tauHI}).
\label{fig_C4.HI.tauhi.ts}}  
\end{figure*}

Figure \ref{fig_C4.HI.tauhi.ts} shows the distribution of $N _{\rm H, cor}$, $T _{\rm S}$, and $\tau _{\rm HI}$. 
For the latter two quantities (which we cannot derive where molecular gas is present), pixels where molecular gas is detected ($^{12}$CO integrated intensity $> 0.5$ K km s$^{-1}$, corresponding to 5 $\sigma$) and/or H$\alpha$ flux is significant, are masked.
$\tau _{\rm HI}$ around the CO clouds is typically $> 1.5$, compared to $\sim 1$ in other areas, suggesting either that H$\,${\sc i} may be somewhat optically thick, or that there is significant CO-dark H$_2$. 
$T _{\rm S}$ ranges from $\sim$ 50--200 K in the region, and is typically $\sim$ 100 K in the vicinity of the CO cloud.
This is higher than found in the Galactic case by \citet{Fukui_etal_2014b, Fukui_etal_2015a}, where optically thick H$\,${\sc i} was found to have $T _{\rm S}$ of 20--40 K and a density of 40--160 cm$^{-3}$.
We suggest that relatively warm and diffuse gas may therefore dominate within the relatively large beam of our data, with the high total $\tau _{\rm HI}$ explained by the long line of sight (i.e. high column density) through the ridge.

\begin{figure*}
 \begin{center}
  \includegraphics[width=2\columnwidth]{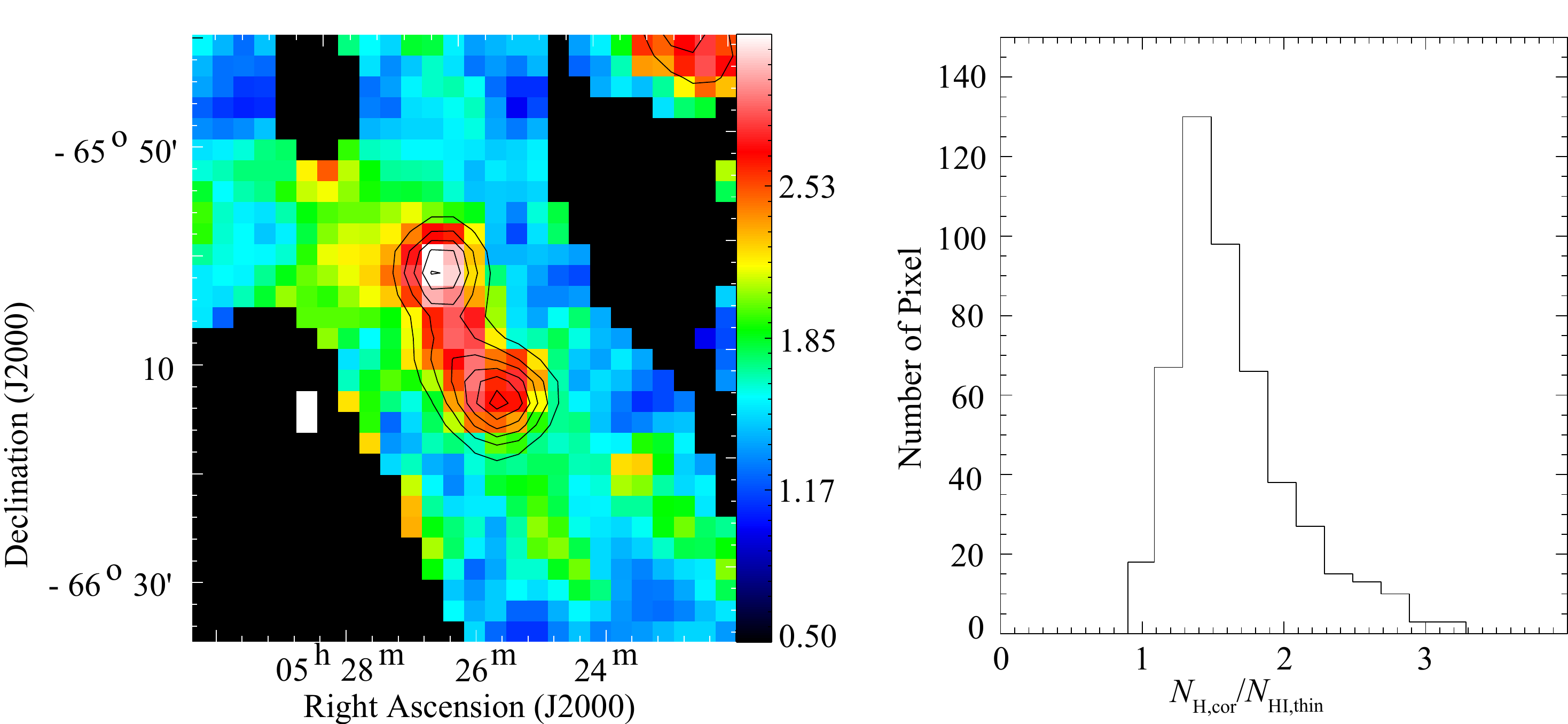}  
 \end{center}
\caption{
(a) Color map of the ratio $N _{\rm H, cor}/N _{\rm HI, thin}$. Contours are NANTEN $^{12}$CO(J=1--0) smoothed to the resolution of the Planck data (5$^{\prime}$). Data that is too faint in H$\,${\sc i} has been masked.
(b) Histogram of the ratio $N _{\rm H, cor}/N _{\rm HI, thin}$.
\label{fig_C4.HI.ratio}}  

\end{figure*}

Figure \ref{fig_C4.HI.ratio}(a) shows the spatial distribution of the ratio $N _{\rm H, cor}/N _{\rm HI, thin}$ within the entire ridge. 
As expected, This is highest ($> 2$) in the CO clouds, with more moderate values throughout the rest of the ridge.
Figure \ref{fig_C4.HI.tauhi.ts}(b) shows a histogram of these ratio values. 
The histogram has a strong peak around 1.4, corresponding to the bulk of the H$\,${\sc i} in the ridge, and shows an extended tail around ratios of 2 to 3, corresponding to the material in the region of the molecular clouds.
This is expected given that $\tau _{353}$ is also tracing the H$_2$ in and around the molecular clouds. 
The total mass derived from $N _{\rm H, cor}$ is $\sim 1\times 10^7$ M$_{\odot}$, which thus in reality corresponds to the sum of both the H$\,${\sc i} and the molecular gas.
Since the GMC mass derived from CO is $\sim1.5 \times$ 10$^6$ M$_{\odot}$ (\citealt{Yamaguchi_etal_2001a}), the remaining 
mass of the ridge is therefore $\sim 8.5 \times 10^6$ M$_{\odot}$ -- roughly 1.7 times greater than the uncorrected (optically thin) H$\,${\sc i} mass.
If CO-dark H$_{2}$ is not dominant, as suggested by \cite{Fukui_etal_2014b, Fukui_etal_2015a}, these results would imply a large amount of ``hidden'' optically thick (i.e. cold) H$\,${\sc i}.
We note that while our corrected mass should likely be regarded as an upper limit, the presence of significant optically thick H$\,${\sc i} in the ridge is broadly consistent with the work of \cite{MarxZimmer_etal_2000}. These authors measured the cold gas fraction in the LMC directly via H$\,${\sc i} absorption measurements against background continuum sources, and reported a high fraction of cool H$\,${\sc i} in the LMC 4 region (cold gas fraction $f _{\rm c} > 60$\%).

\subsection{Filamentary Nature of the H$\,${\sc i} Ridge \label{c4_HIfilament}}

\subsubsection{Identification of Filamentary Features}

The computational identification of H$\,${\sc i} filaments is not straightforward. 
This is because the H$\,${\sc i} gas contains a wide spread of temperatures and densities, which tend to cluster into two distinct phases -- dense, cold gas (cold neutral medium; CNM), and warm, diffuse gas (warm neutral medium; WNM).
The warm component has a high temperature (typically several thousand K), such that 
even diffuse, optically thin gas ($\tau < 0.1$) readily reaches brightness temperatures, $T_\mathrm{b}$, of at least several 10 K.  
On the other hand, the densest, coldest material, which is optically thick and which one might expect to be the main component of an H$\,${\sc i} filament, is of comparable brightness, despite often significantly higher column densities (since $T_\mathrm{b}$ cannot exceed $T_\mathrm{S}$). 
Due to these reasons, the contrast of the observed H$\,${\sc i} brightness temperature distribution is low -- only poorly reflecting complex column-density distribution of the two phases.

While a number of algorithmic approaches to ISM filament finding exist \citep[e.g.][]{Arzoumanian_etal_2011, Clark_etal_2014, Koch_Rosolowsky_2015}, the difficulties described above lead us to opt instead for a core identification algorithm.
The basic concept is as follows:
First, the core identification algorithm (see below) identifies the positions of local peaks of each velocity channel map. Filaments may then be defined by eye, by connecting cores along contour ridges. 
(Note that this is qualitatively similar to the approach taken by algorithms such as DisPerSE; \citealt{Sousbie_2011}.)
The physical parameters of the resultant filaments, such as width, length, and line mass, may then be derived from those of the chained cores.
This identification method may provide a rough estimate of the distribution and physical parameters of the H$\,${\sc i} filaments.
We refer to the identified structures as ``{\it filamentary features}''.

For the core identification process, we adopted the dendrogram method of \cite{Rosolowsky_etal_2008}.
Structural analyses using dendrograms have previously been performed to investigate hierarchical, cloud-to-core gas structures in position-position-velocity (PPV) space (e.g., \citealt{Goodman_etal_2009, Kauffmann_etal_2013}), and the method is also useful for identifying filamentary structure in dense clouds by selecting cores with high aspect ratios (e.g., \citealt{Lee_etal_2014, Storm_etal_2014}). 

The dendrogram method provides a tree diagram that characterizes how and where structures surrounding local maxima in the PPV space merge. 
Structures increase their PPV volume from the local maxima with decreasing intensity level until they encounter adjacent structures. 
Local maxima in intensity are the ``leaves'' of the tree, which are the finest structure identified by the method. 
Two thresholds are set for the core identification criteria, one is ``minimum value'' that is simply to get rid of any structure peaking below this minimum, and the other is ``minimum delta'' that defines a minimum height required for a structure to be retained and is intended to avoid identifying noise as local maxima.
The merge level, defined as the iso-contour which encircles two or more leaves, creates ``branch''. 
Leaves and branches may merge into a lower level branch which can repeat the cycle of growth and merger until the minimum value of the flux density is reached.

For our H$\,${\sc i} data, all ``leaves'' tend to be included in a single large ``branch'' since each fine structure is rather buried in the ambient component. 
It is impossible to define filaments by identifying leaves or branches with high aspect ratios. 

The detailed procedure for identifying filamentary features is as follows:
\begin{enumerate}
\item A new set of H$\,${\sc i} channel maps are generated, each integrated over a velocity interval of 10 km s$^{-1}$ (the characteristic line width of a typical H$\,${\sc i}  component), with a velocity step of 5 km s$^{-1}$. The central velocity of the first map is 270.4 km s$^{-1}$ . 
\item The dendrogram method is used to identify a population of cores in each channel map. 
The minimum value is set to 78 [K km s$^{-1}$] (corresponding to the 5$\sigma$ noise level in integrated intensity), and the minimum delta is set to 93 [K km s$^{-1}$] (6$\sigma$ noise level) for the 285.4 to 300.4 km s$^{-1}$ channel maps, and 78 [K km s$^{-1}$] for the others.
\item Cores are connected by eye into filamentary features. 
Collections of cores that can be merged within elongated H$\,${\sc i} contours are connected by drawing a line that follows the direction of elongation, and passes through the central positions of each core.
If the enclosing contours are too diffuse (typically wider than twice the size of the cores), these are excluded from the identification.
\item Highly elongated structures seen in the contour maps without clear strings of cores are also identified as filamentary features. If any core is found along the filamentary feature, it is included as part of the feature.
\item A filamentary feature is permitted to extend elongate outside the boundaries of the cores at either end, for as long as the contours appear to trace a coherent part of the filament.
\item If similar filamentary features are seen in 2 or more channels, a single channel is selected in which it appears the most prominent, in order to avoid duplication. 
\item If the velocity gradient of a filamentary feature is larger than the integration range of a single channel map (10 km s$^{-1}$), it must be identified in 3 or more channels. 
If filamentary features continue to elongate in more than 3 channels, these are merged into one.
\end{enumerate}

Note that contiguous structures in velocity channel maps may not necessarily reflect material that is truly contiguous in 3-D space. 
Indeed, since the line widths of the H$\,${\sc i} emission are quite large, contamination from spatially unrelated components may be significant. 
However, since we are unable to identify such contamination, we henceforth assume that all contiguous structures in velocity channel maps are true filamentary features.

Note also that we have adopted 5 sigma noise level for the two parameters, in order to pick up as many core-like features as possible, while suppressing the false identification of noise peaks. With lower thresholds, we found that more noise-like features were identified as cores, and with higher ones, we found that individual cores tended to be merged to single large structures; both are not desirable here.
See for example the 275.4 km s$^{-1}$ panel: noise identification is effectively avoided, while prominent cores are picked up sufficiently.
The higher minimum delta adopted in the 285.4 to 300.4 km s$^{-1}$ panels (93 [K km s$^{-1}$], corresponding to 6 sigma) was chosen for pragmatic reasons: at 5 sigma a lot of small noise-like cores were identified in these panels, especially in areas of high H$\,${\sc i} luminosity. 

Figures \ref{fig_C4.HI.fil.01} to \ref{fig_C4.HI.fil.03} show the results of this identification procedure.
Filamentary features were identified in all channels. 
In the 280.4 km s$^{-1}$ and 285.4 km s$^{-1}$ panels, clear spatial correlations can be seen between the CO and filamentary features, possibly suggesting that the filamentary nature of the H$\,${\sc i} is key to their formation process.
Prominent CO emission is also seen in the 290.4 km s$^{-1}$ and 295.4 km s$^{-1}$ panels, which cover the peak velocities of typical H$\,${\sc i} spectra in the ridge. In this velocity range, the H$\,${\sc i} gas is likely becoming optically thick (see Fig. \ref{fig_C4.HI.tauhi.ts}; $\tau _{\rm HI}$ is typically $\gtrsim1$ in this region), meaning that structures may be buried in the smooth intensity distribution. Likely for this reason, no filamentary features were identified within the bright centre of the ridge.
However, filamentary features are seen in the lower brightness areas surrounding it, 
suggesting that a similar morphology may indeed persist in the ridge centre, and is merely obscured due to the high optical depth of the gas.
Similarly, in the extreme velocity channels where optical depths are presumably lower (270.4 km s$^{-1}$ to 280.4 km s$^{-1}$, and 300.4 km s$^{-1}$ to 315.4 km s$^{-1}$), many filamentary features are found.
The number of the features identified is greater at the red end of the channel maps (300.4 km s$^{-1}$ to 315.4 km s$^{-1}$), than the blue end (270.4 km s$^{-1}$ to 280.4 km s$^{-1}$), reflecting the presence of extended ambient gas on the LMC 5 side at lower velocities.

In total, 39 filamentary features are identified.
These are overlaid on a Herschel 500 $\mu$m dust map in figure \ref{fig_C4.HI.fil.all}, color-coded by velocity range. 
The distribution of the filamentary features is complicated.
Several roughly follow the dust intensity distribution, most are oriented along the direction of the ridge, although several are perpendicular, and   
their center of curvature is more commonly oriented towards LMC 5 than LMC 4.

\begin{figure*}
 \begin{center}
  \includegraphics[width=1.6\columnwidth]{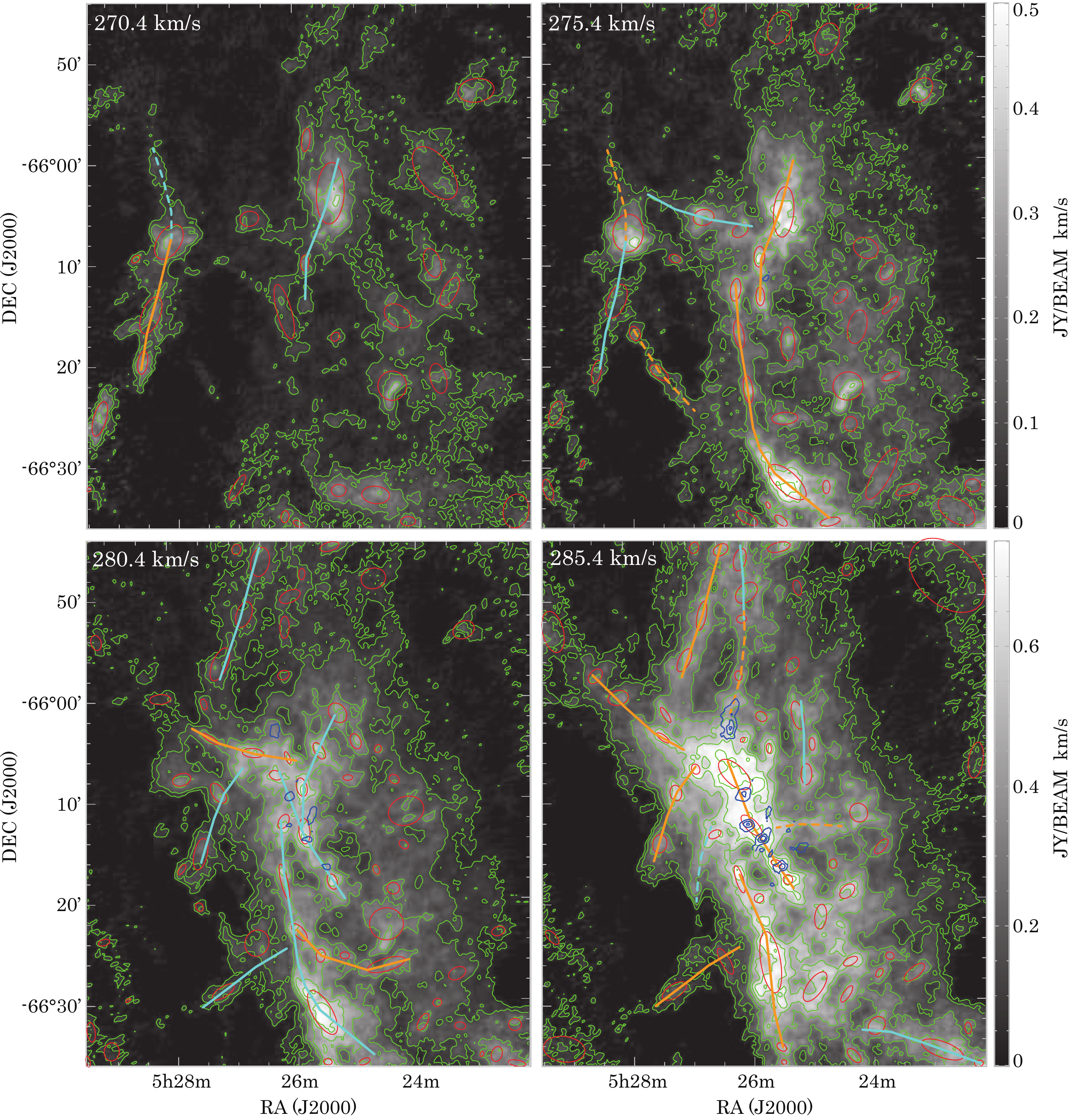}  
 \end{center}
\caption{
Channel maps used for the identification of filamentary features. Channels with central velocities of 270.4 km s$^{-1}$ to 285.4 km s$^{-1}$ are shown, each with a velocity width of 10 km s$^{-1}$. The greyscale image and green contours show H$\,${\sc i} integrated intensity. Contours begin at the 5$\sigma$ level, and are incremented in steps of 10$\sigma$ (0.065 and 0.13 Jy/Beam km s$^{-1}$). Blue contours are ASTE $^{12}$CO($J$=3--2), starting from 4 K km s$^{-1}$ and incrementing in steps of 8 K km s$^{-1}$. Red contours show the cores identified from the dendrogram analysis. Orange lines indicate features identified by connecting cores, orange dashed lines are those identified from the H$\,${\sc i} contour distribution alone, and cyan lines indicate features identified in adjacent channels (see point 6 of the identification procedure). 
\label{fig_C4.HI.fil.01}}  

\end{figure*}

\begin{figure*}
 \begin{center}
  \includegraphics[width=1.6\columnwidth]{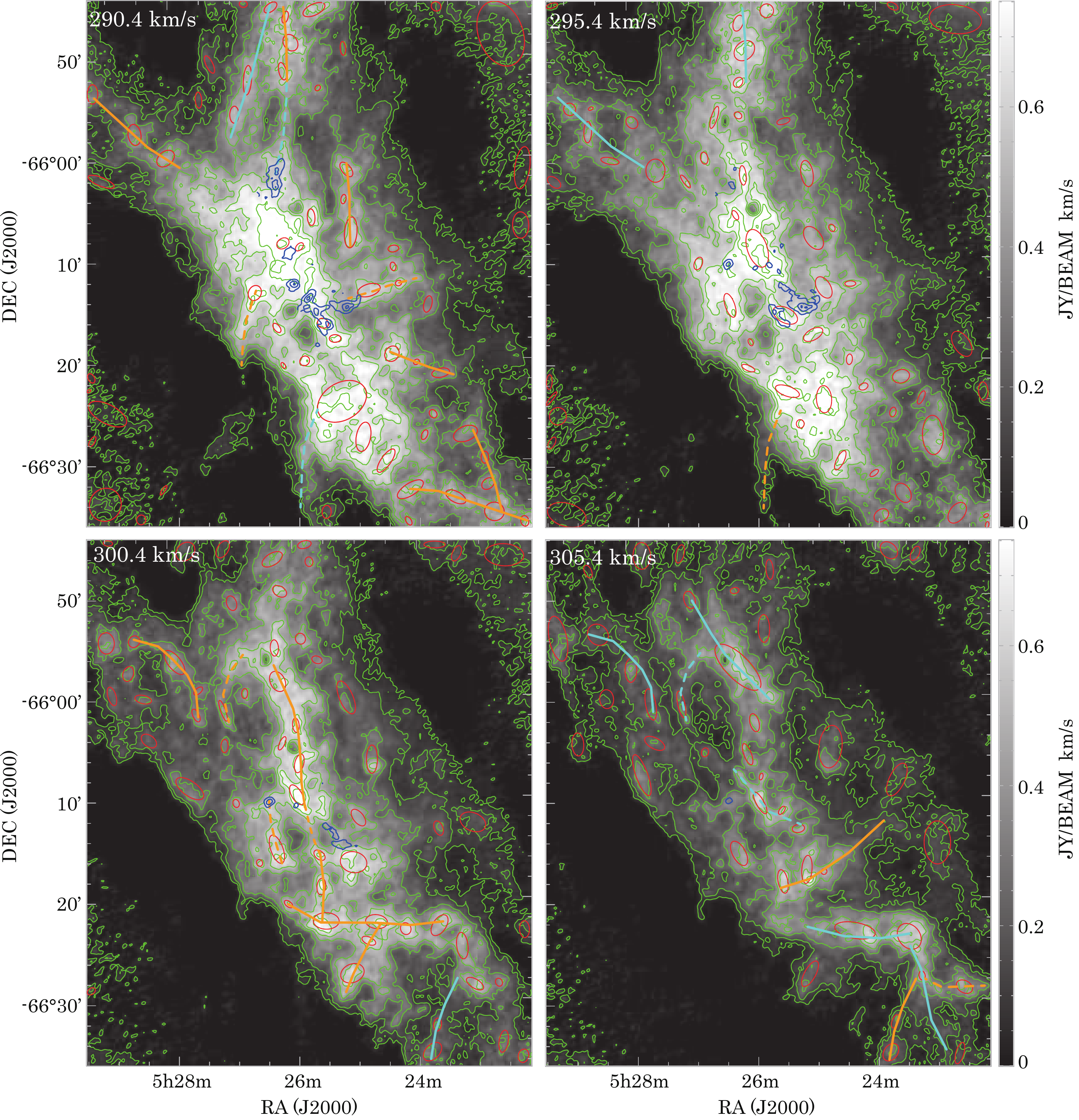}  
 \end{center}
\caption{
Continued channel maps for the identification of filamentary features. Channels of 290.4 km/s to 305.4 km/s are shown.
\label{fig_C4.HI.fil.02}}  

\end{figure*}

\begin{figure*}
 \begin{center}
  \includegraphics[width=1.6\columnwidth]{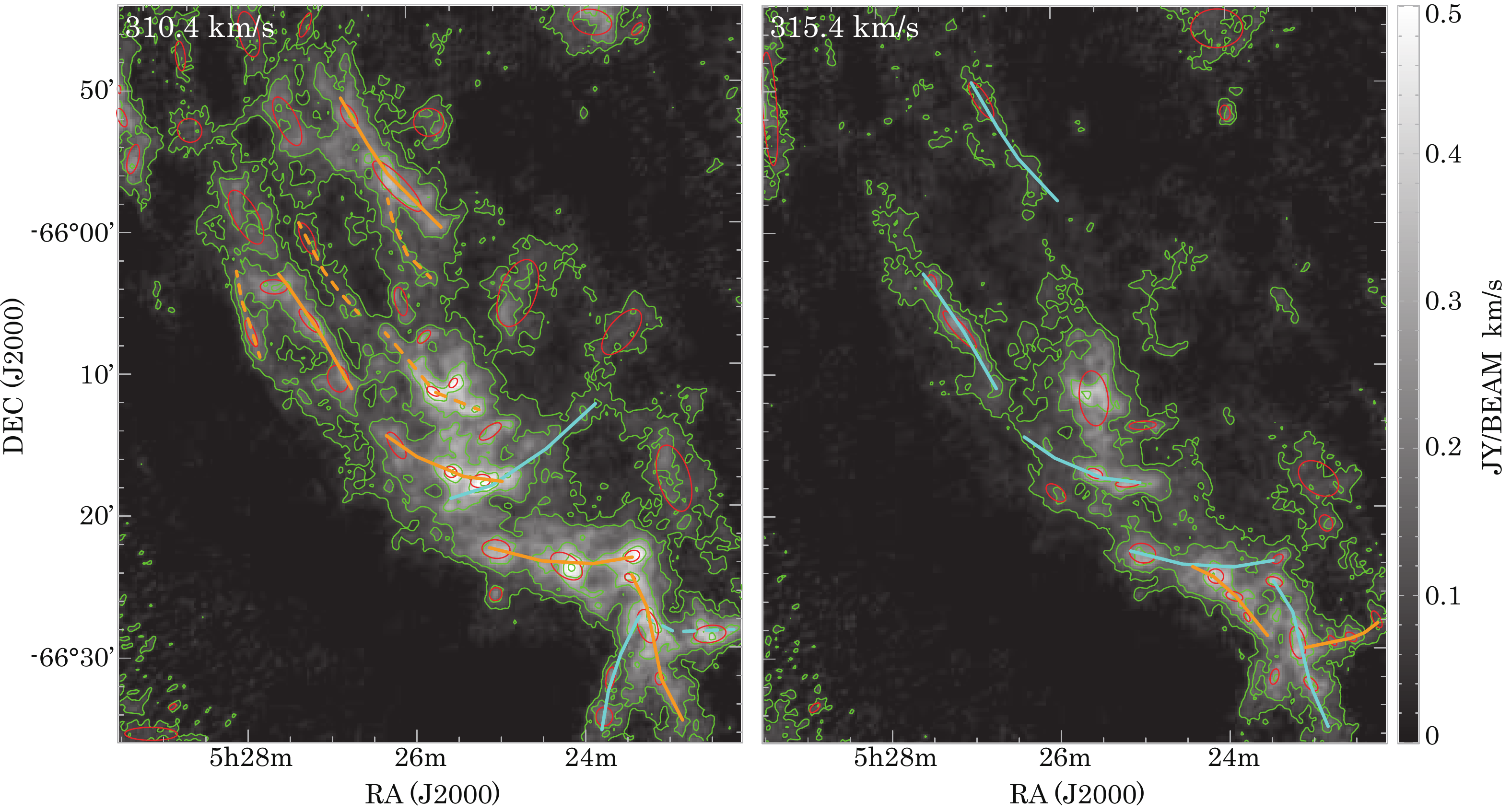}  
 \end{center}
\caption{
Continued channel maps for the identification of filamentary features. Channels of 310.4 km/s and 315.4 km/s are shown.
\label{fig_C4.HI.fil.03}}  

\end{figure*}

\begin{figure}
 \begin{center}
  \includegraphics[width=\columnwidth]{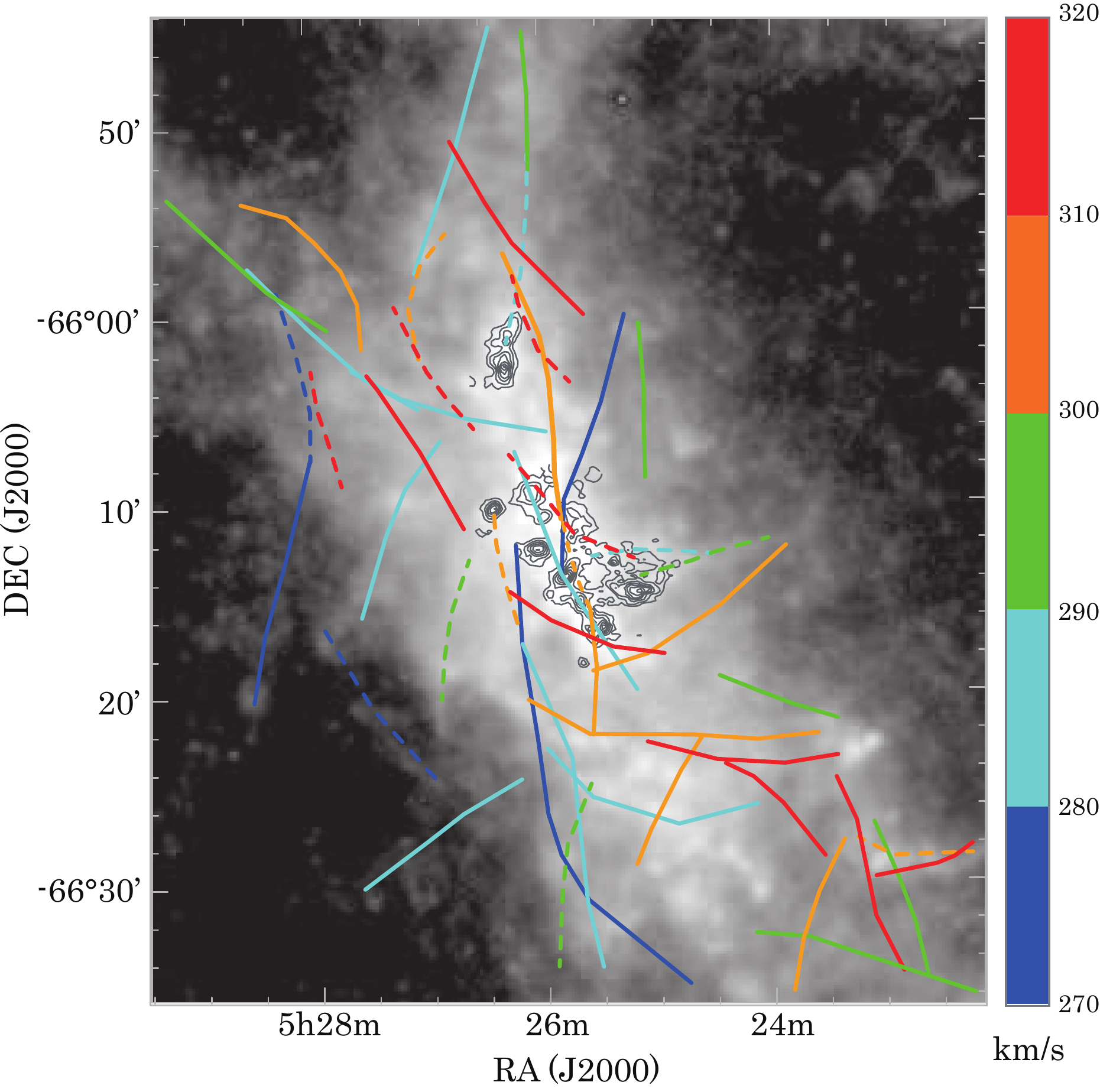}  
 \end{center}
\caption{
The composition map of the all identified filamentary features overlaid on the Herschel 500 $\mu$m image (\citealt{Meixner_etal_2013}). The filamentary features are colored by their velocity range.
\label{fig_C4.HI.fil.all}}  

\end{figure}

\subsubsection{Physical Properties of the Filamentary Features \label{s4_filpara}}

\begin{figure*}
 \begin{center}
  \includegraphics[width=2\columnwidth]{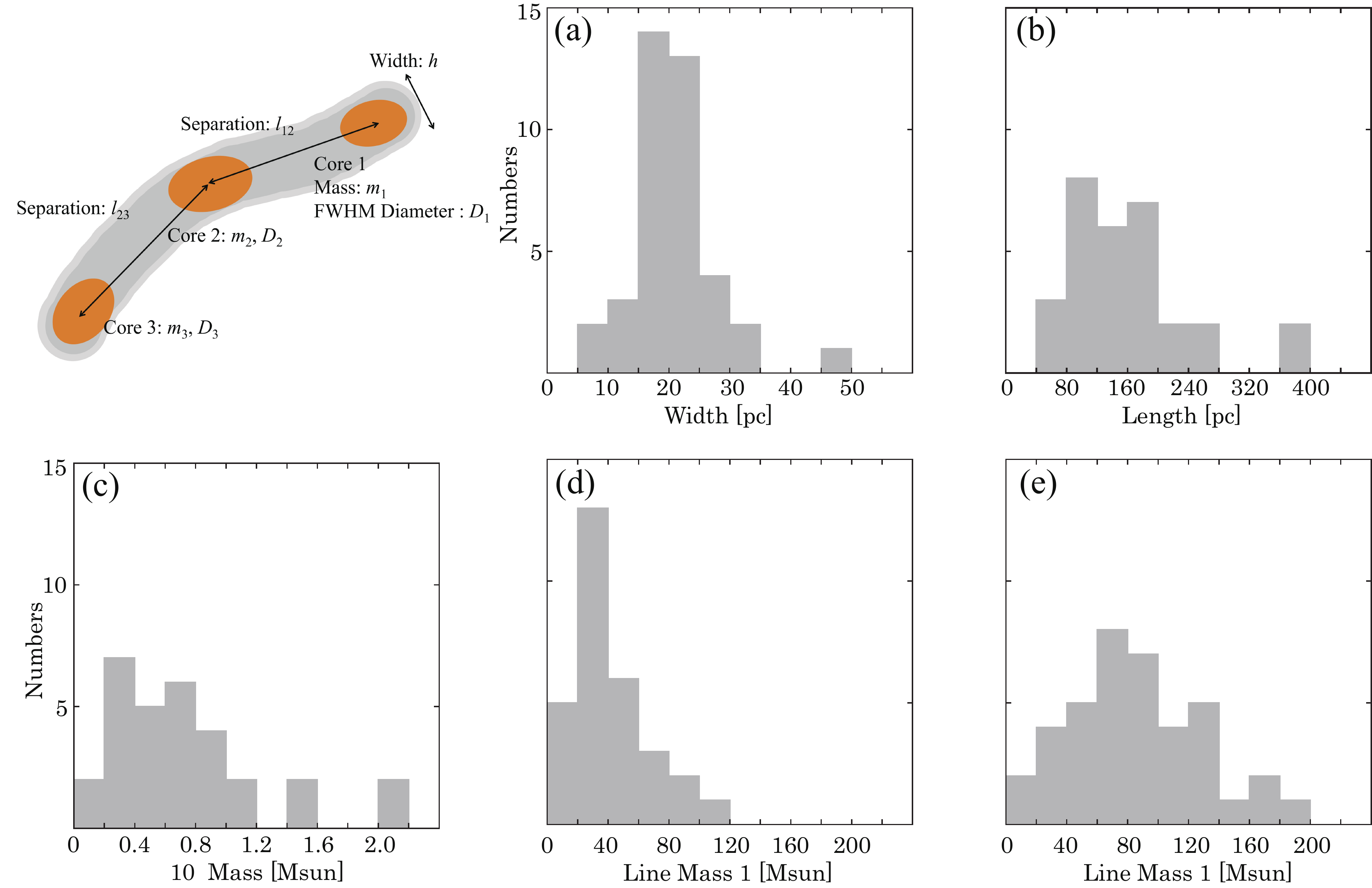}  
 \end{center}
\caption{
Histograms of the physical parameters of the filamentary features. The top-left pannel is a schematic view of the physical properties of the filamentary feature. 
(a) Width of the filamentary features defined by Equation \ref{eq_filwidth}, (b) length of the filamentary features defined by Equation \ref{eq_fillength}, (c) total mass of the filamentary features defined by Equation \ref{eq_filmass}, (d) and (e) line mass of the filamentary features defined by Equations \ref{eq_linemass1} and \ref{eq_linemass2}.
\label{fig_C4.HI.fil.hist}}  
\end{figure*}

We may estimate the physical properties of the filamentary features, such as width, length, mass, and line mass, from the properties of the H$\,${\sc i} cores that compose them. 
The top left panel of Figure \ref{fig_C4.HI.fil.hist} shows a schematic view of the physical properties of a filamentary feature and its constituent H$\,${\sc i} cores. Following the notation in this figure, the physical parameters of the filamentary features are estimated as described below.

The width of a filamentary feature, $h$, may be roughly estimated from the mean value of the FWHM diameter of its H$\,${\sc i} cores:
\begin{equation}
h \sim \frac{1}{N}\sum _{i=1}^{N} D _{i},
\label{eq_filwidth}
\end{equation}
where $N$ is the number of cores. 
Each FWHM diameter $D _{i}$ is given by the geometric mean of the dispersion of the semi-major axis $\sigma _{x, i}$ and semi-minor axis $\sigma _{y, i}$ (as output from the dendrogram software),
\begin{equation}
D _{i} = \sqrt{8 \ln 2 \cdot \sigma _{x, i} \sigma _{y, i}}.
\end{equation} 

The precise determination of the length of the filamentary feature, $L$, is difficult, because the definition of the end points is quite loose.
One reasonable estimation can be given by using the separation between the H$\,${\sc i} cores $l _{ij}$,
\begin{equation}
L \sim \sum _{i<j}^{N} l _{ij} + h,
\label{eq_fillength}
\end{equation}
where the width, $h$, is added to approximate the extent of the end core past their central positions. 
This equation is applicable to filamentary features identified by connecting cores

The total mass of a filamentary feature, $M$, is also difficult to estimate, because its true extent in PPV space is not well determined.
One simple approach is to simply sum up the mass of the H$\,${\sc i} cores, as estimated under the optically thin assumption,
\begin{equation}
M > \sum _{i=1}^{N} m _{i},
\label{eq_filmass}
\end{equation}
which gives an approximate lower limit. 
Note that H$\,${\sc i} core masses are simply derived from the emission in a single channel map with velocity width of 10 km s$^{-1}$, meaning that some fraction of the relevant velocity component may be excluded, while there may also be contamination from unrelated velocity components. 
Since the actual extent of each component in velocity space cannot be determined (due to significant spectral overlap), this uncertainty is inevitable. 
We also note that high H$\,${\sc i} optical depths are another reason that the actual mass of the H$\,${\sc i} may be higher than derived here.

The line mass, $M _{L}$, can be estimated in two ways.
The first is to divide the total mass of a filamentary feature, $M$, by its length, $L$,
\begin{equation}
M _{L} \sim M/L.
\label{eq_linemass1},
\end{equation}
which gives an approximate lower limit to the line mass. 
An alternative is to estimate the line mass only in the regions of the H$\,${\sc i} cores, by taking the average of value the masses of the cores divided by their FWHM diameters.
\begin{equation}
M _{L} \sim \frac{1}{N} \sum _{i}^{N} \frac{m _{i}}{D _{i}}.
\label{eq_linemass2}
\end{equation}
This gives a higher estimate of the line mass, since it is calculated only in what we assume to be the densest parts of the filamentary features.
Here, the latter expression is adopted.

The derived parameters are summarized in Table \ref{fig_C4.HI.fil.hist}, and histograms are shown in Figure \ref{fig_C4.HI.fil.hist}.
The median (minimum--maximum) of each parameter is $h = 21$ pc (8--49 pc), $L = 118$ pc (11--400 pc), $M =$ 6,200 M$_{\odot}$ (1,600--21,000 M$_{\odot}$) and $M _{L} = 90$ M$_{\odot}$/pc (20--190 M$_{\odot}$/pc).
The most massive and prominent filamentary feature No. 11 ($M =$ 21,000 M$_{\odot}$, $M _{L} = 190$ M$_{\odot}$/pc) is associated with the molecular clumps.

The number distribution of the width, $h$, has a strong peak around 20 pc, with a standard deviation of 7 pc. 
Since previous H$\;${\sc i} survey data (60$^{\prime \prime}$, $\sim 15$ pc) has been barely able to resolve these size-scales, the discovery of this characteristic size-scale for H$\;${\sc i} structure 
is a key result of our high-resolution observations.
The diameter of the molecular clumps is $\sim$ 10 pc, which is somewhat smaller than this estimated 20 pc filamentary feature width.
The length and the mass of the filamentary features show broad peaks around the $100\sim200$ pc and $0.3\sim1.0\times 10^4$ M$_{\odot}$ ranges, respectively. 
Under the second definition of line mass, the distribution shows a broad peak around $M _{L} \sim 90$ M$_{\odot}$ pc$^{-1}$, corresponding to an average density of 10 cm$^{-3}$ for a cylinder of 20 pc width.
This is one order of magnitude denser than expected for WNM ($\lesssim 1$ cm$^{-3}$), which implies that the filamentary features contain large amounts of cooler, denser gas, consistent with the results of section \ref{s4_opacity}. We may speculate that this high CNM fraction and filamentary structure are both a result of shock-compression by the SGSs, a scenario that is discussed in more detail below.

An approximate upper limit on the density under the assumptions above is $\sim$ 20 cm$^{-3}$, which arises from the maximum measured line mass of $M _{L} \sim 190$ M$_{\odot}$ pc$^{-1}$.
However, given that H$\,${\sc i} column density increases by a factor of $\sim1.4$--2.0 when opacity is taken into account (see \S\ref{s4_opacity}),
the average density of the filamentary features may be in fact be higher, roughly in the range $\sim10$--40 cm$^{-3}$.
We note that this average density is still lower than that of the optically thick H$\;${\sc i} found by \citet{Fukui_etal_2014b, Fukui_etal_2015b} in the Galactic case (40--160 cm$^{-3}$), supporting the idea that the filamentary features contain both cold and warm components. 
Yet for a purely cold filamentary feature with a width of 20 pc, a spin temperature of 100 K, a velocity width of 10 km s$^{-1}$, and a density range of 10--40 cm$^{-3}$, the mean $\tau _{\rm HI}$ over the line FWHM (see Equation \ref{eq_tauHI}) falls in the range 0.3--1.4, which is still lower than the total optical depth through the ridge as estimated in section \ref{s4_opacity} ($\tau _{\rm HI} \gtrsim$ 1.5).
This is consistent with a picture in which multiple filamentary features do indeed overlap along the line of sight, as suggested by their arrangement in Figure \ref{fig_C4.HI.fil.all}.

The theoretical critical line mass for an isothermal, self-gravitating cylinder with no magnetic support (\citealt{Ostriker_1964}) is given by
\begin{eqnarray}
M _{\rm L, crit} &=& 2 c_{\rm s}^{2} /G \\
				& \sim & 1.67 \times \left( T \; [{\rm K}]  \right) \; [{\rm M} _{\odot}\;{\rm pc}^{-1}].
\end{eqnarray}
For the estimated harmonic mean spin temperature of the H$\,${\sc i} ridge (100--200 K), the critical line mass is typically 200--300 M$_{\odot}$/pc.
This is higher than the typical line mass estimated above (20--190 M$_{\odot}$/pc), 
even for the most massive one (No. 11, the only feature clearly associated with molecular gas). 
Even applying an opacity correction factor of 1.4--2.0 does not bring the estimated line mass up to the critical values, strongly implying that these filamentary features are not self-gravitating structures. 
This is consistent with the fact that most show no evidence of associated molecular clumps,  
suggesting that that an increase in density and/or mass is required before molecular gas and star formation can occur.

\begin{table*}
 \caption{Parameters of filamentary features}
 \label{c4_fil_para_1}
 \begin{center}
 \scalebox{0.8}{
  \begin{tabular}{lcllcccccccccc}
   \hline \hline
Channel &ID &\multicolumn{2}{c}{Position of the Cores}&&\multicolumn{3}{c}{Core parameters}&&\multicolumn{5}{c}{Filament Parameters}\\
 \cline{3-4} \cline{6-8} \cline{10-14}
Velocity&Number&R.A.($J$2000)&Dec($J$2000)&&axis $\sigma _{x}$$\times$$\sigma _{y}$&Diameter&Mass&&Width&Length&Mass&Line Mass 1&Line Mass 2\\
$[$km s$^{-1}]$&&h:m:s&d:$^{\prime}$:$^{\prime \prime}$&&[pc$\times$pc]&[pc]&[M$_{\odot}$]&&[pc]&[pc]&[10$^4$ M$_{\odot}$]&[M$_{\odot}$/pc]&[M$_{\odot}$/pc]\\
\hline
270.4&1&5:28:32.2&-66:19:41.6&&34$\times$14&22&1100&&34&180&0.74&40&58\\
&&5:28:23.2&-66:15:2&&56$\times$26&38&1700&&&&&&\\
&&5:28:3.6&-66:7:48.7&&56$\times$33&43&3500&&&&&&\\
\hline
275.4&2&5:24:44.1&-66:35:57.8&&30$\times$11&18&1100&&26&370&1.5&57&90\\
&&5:25:26.5&-66:32:0.4&&69$\times$30&45&9700&&&&&&\\
&&5:26:4&-66:22:45.5&&40$\times$13&23&1300&&&&&&\\
&&5:26:12.1&-66:17:18.5&&40$\times$14&24&1200&&&&&&\\
&&5:26:14.7&-66:12:53.2&&33$\times$14&22&1500&&&&&&\\
&3&5:25:50.7&-66:13:34.9&&27$\times$12&18&1200&&28&130&1.1&72&110\\
&&5:25:50.8&-66:9:38.9&&30$\times$15&21&1500&&&&&&\\
&&5:25:26.5&-66:5:14.9&&73$\times$29&46&8700&&&&&&\\
&4&5:27:31.6&-66:20:56.7&&32$\times$15&22&520&&20&&&&23\\
&&5:27:54.7&-66:16:58.4&&21$\times$15&18&390&&&&&&\\
&5&5:27:59.1&-66:7:0&&54$\times$44&49&6400&&49&&&&130\\
\hline
280.4&6&5:25:24.2&-66:18:16.9&&23$\times$14&18&1700&&21&150&0.83&92&130\\
&&5:25:50.9&-66:13:14.9&&37$\times$16&24&3900&&&&&&\\
&&5:25:53&-66:8:52.9&&27$\times$15&20&2700&&&&&&\\
&7&5:24:31.8&-66:26:33.6&&56$\times$18&32&2900&&23&140&0.6&60&88\\
&&5:25:36.7&-66:25:35.2&&19$\times$13&16&1200&&&&&&\\
&&5:25:57.5&-66:23:14.6&&31$\times$14&21&1900&&&&&&\\
\hline
285.4&8&5:25:32.9&-66:34:39.2&&16$\times$9.9&12&860&&28&250&2.1&100&180\\
&&5:25:42.8&-66:26:14.9&&89$\times$28&50&17000&&&&&&\\
&&5:26:12.3&-66:17:59.6&&41$\times$10&21&2600&&&&&&\\
&9&5:27:2.4&-65:56:9.9&&36$\times$16&24&2300&&22&170&0.66&67&99\\
&&5:26:44.8&-65:51:46.8&&43$\times$13&23&2600&&&&&&\\
&&5:26:25.2&-65:45:16.8&&23$\times$16&19&1700&&&&&&\\
&10&5:27:30&-66:14:18.4&&27$\times$20&23&2000&&19&120&0.57&89&100\\
&&5:27:13.3&-66:9:19.7&&23$\times$17&20&2300&&&&&&\\
&&5:26:53.5&-66:7:0&&15$\times$12&13&1400&&&&&&\\

\hline
  \end{tabular}
  } 
 \end{center}
  \footnotetext{}{ Col.(1): The central velocity of the channel in which the filamentary feature is identified. Col.(2): Serial numbers of the filamentary features. Col.(3)(4): Central positions of H$\,${\sc i} cores that are used for the identification of the filamentary features. Col.(5)(6)(7): Major and minor axis, geometric mean diameter of the equivalent ellipse, and corresponding mass of the H$\,${\sc i} cores. Col.(8)(9)(10)(11)(12): Parameters of the filamentary features. Corresponding equations are (\ref{eq_filwidth}), (\ref{eq_fillength}), (\ref{eq_filmass}), (\ref{eq_linemass1}), and (\ref{eq_linemass2}). For filamentary features that are identified by the contour distribution alone, length, mass, and line mass 1 cannot be derived, and empty values are listed here.}
\end{table*}

\begin{table*}
 \caption{Filament Parameters}
 \label{c4_fil_para_2}
 \begin{center}
 \scalebox{0.8}{
  \begin{tabular}{lcllcccccccccc}
   \hline \hline
Channel &ID &\multicolumn{2}{c}{Position of the Cores}&&\multicolumn{3}{c}{Core parameters}&&\multicolumn{5}{c}{Filament Parameters}\\
 \cline{3-4} \cline{6-8} \cline{10-14}
Velocity&Number&R.A.($J$2000)&Dec($J$2000)&&axis $\sigma _{x}$$\times$$\sigma _{y}$&Diameter&Mass&&Width&Length&Mass&Line Mass 1&Line Mass 2\\
$[$km s$^{-1}]$&&h:m:s&d:$^{\prime}$:$^{\prime \prime}$&&[pc$\times$pc]&[pc]&[M$_{\odot}$]&&[pc]&[pc]&[10$^4$ M$_{\odot}$]&[M$_{\odot}$/pc]&[M$_{\odot}$/pc]\\
\hline 
&11&5:25:24.4&-66:18:6&&14$\times$14&14&1800&&23&170&2.1&140&190\\
&&5:25:35.8&-66:16:27.5&&16$\times$9.1&12&1300&&&&&&\\
&&5:25:59.5&-66:12:41.1&&39$\times$17&26&5500&&&&&&\\
&&5:26:15.4&-66:7:37.1&&57$\times$28&40&13000&&&&&&\\
&12&5:27:18.8&-66:28:53.5&&40$\times$19&27&1200&&25&85&0.2&24&38\\
&&5:26:27.5&-66:26:3.7&&42$\times$13&23&710&&&&&&\\
&13&5:26:6.3&-66:5:55.8&&19$\times$6.5&11&700&&19&340&0.75&46&67\\
&&5:26:40.2&-66:5:23.9&&31$\times$13&20&2000&&&&&&\\
&&5:27:21.4&-66:3:55.6&&27$\times$7.5&14&710&&&&&&\\
&&5:28:7.5&-65:59:44.1&&30$\times$21&25&1700&&&&&&\\
&&5:28:29.4&-65:57:31.4&&25$\times$16&20&1100&&&&&&\\
&&5:29:14.2&-65:53:14&&28$\times$15&21&1400&&&&&&\\
\hline
290.4&14&5:22:46.1&-66:35:10.4&&28$\times$26&27&3100&&21&120&0.76&70&81\\
&&5:22:52.5&-66:32:26.4&&20$\times$15&17&1200&&&&&&\\
&&5:22:55.5&-66:29:50.3&&17$\times$11&14&570&&&&&&\\
&&5:23:14.6&-66:27:29.3&&36$\times$21&28&2800&&&&&&\\
&15&5:22:16&-66:36:7.1&&15$\times$11&13&980&&23&170&1&83&110\\
&&5:22:46.1&-66:35:10.4&&28$\times$26&27&3100&&&&&&\\
&&5:23:24.5&-66:33:43.5&&32$\times$16&22&2500&&&&&&\\
&&5:24:9.6&-66:32:57.3&&43$\times$18&27&3400&&&&&&\\
&16&5:25:5.9&-66:7:30.5&&45$\times$18&28&4700&&19&92&0.71&74&110\\
&&5:25:12.6&-66:4:11.9&&14$\times$7.4&10&590&&&&&&\\
&&5:25:9.4&-66:1:16.2&&25$\times$13&18&1800&&&&&&\\
&17&5:26:10&-65:51:25.7&&26$\times$11&16&920&&17&80&0.42&60&78\\
&&5:26:1.8&-65:48:37.5&&25$\times$20&22&2400&&&&&&\\
&&5:26:6.6&-65:46:5.3&&18$\times$8.4&12&890&&&&&&\\
&18&5:24:48.8&-66:13:14.3&&34$\times$17&24&3400&&16&44&0.39&79&100\\
&&5:24:19.6&-66:12:34.2&&9.3$\times$7.8&8.5&530&&&&&&\\
&19&5:26:40.6&-66:13:19&&22$\times$18&20&2900&&20&&&&140\\
\hline
295.4&& no data &&&&&&&&&&&\\
\hline
300.4&20&5:23:40&-66:22:48.2&&34$\times$19&25&3000&&21&220&1.5&100&130\\
&&5:24:13.4&-66:23:6.3&&17$\times$15&16&1900&&&&&&\\
&&5:24:42.2&-66:22:23.5&&32$\times$20&25&3600&&&&&&\\
&&5:25:31.9&-66:22:14.3&&37$\times$24&30&6100&&&&&&\\
&&5:26:8.6&-66:20:31.6&&14$\times$7.7&11&680&&&&&&\\

\hline
  \end{tabular}
  } 
 \end{center}
  \footnotetext{}{(Continued Table) 
	}
\end{table*}

\begin{table*}
 \caption{Filament Parameters}
 \label{c4_fil_para_3}
 \begin{center}
 \scalebox{0.8}{
  \begin{tabular}{lcllcccccccccc}
   \hline \hline
Channel &ID &\multicolumn{2}{c}{Position of the Cores}&&\multicolumn{3}{c}{Core parameters}&&\multicolumn{5}{c}{Filament Parameters}\\
 \cline{3-4} \cline{6-8} \cline{10-14}
Velocity&Number&R.A.($J$2000)&Dec($J$2000)&&axis $\sigma _{x}$$\times$$\sigma _{y}$&Diameter&Mass&&Width&Length&Mass&Line Mass 1&Line Mass 2\\
$[$km s$^{-1}]$&&h:m:s&d:$^{\prime}$:$^{\prime \prime}$&&[pc$\times$pc]&[pc]&[M$_{\odot}$]&&[pc]&[pc]&[10$^4$ M$_{\odot}$]&[M$_{\odot}$/pc]&[M$_{\odot}$/pc]\\
\hline 
&21&5:25:31.9&-66:22:14.3&&37$\times$24&30&6100&&21&95&0.99&110&150\\
&&5:25:36.2&-66:18:50.7&&24$\times$13&17&2200&&&&&&\\
&&5:25:37.6&-66:15:44.4&&19$\times$12&15&1600&&&&&&\\
&22&5:25:58.9&-66:9:43.8&&32$\times$14&22&4100&&19&220&1.2&83&120\\
&&5:25:55.9&-66:6:38.8&&27$\times$13&19&1700&&&&&&\\
&&5:25:58.8&-66:2:51&&21$\times$9.5&14&1400&&&&&&\\
&&5:26:15.7&-65:58:28.5&&31$\times$21&25&2400&&&&&&\\
&&5:26:18.4&-65:55:8.9&&24$\times$12&17&1800&&&&&&\\
&23&5:27:35.6&-66:1:40.2&&20$\times$9&14&740&&17&150&0.51&47&65\\
&&5:27:33.1&-65:59:14.6&&16$\times$7.9&11&490&&&&&&\\
&&5:27:55.7&-65:56:22.3&&42$\times$18&28&3100&&&&&&\\
&&5:28:34.2&-65:54:11.4&&17$\times$15&16&850&&&&&&\\
&24&5:25:6.5&-66:27:27.3&&35$\times$26&30&4600&&24&82&0.98&110&130\\
&&5:24:49.9&-66:24:16&&21$\times$11&15&1600&&&&&&\\
&&5:24:42.2&-66:22:23.5&&32$\times$20&25&3600&&&&&&\\
&25&5:26:19.3&-66:15:5.3&&43$\times$17&27&4600&&27&&&&170\\
&26&5:27:9.3&-66:1:15.6&&32$\times$7.1&15&1100&&15&&&&71\\
\hline
305.4&27&5:23:48.3&-66:35:17.9&&28$\times$20&23&1400&&16&120&0.26&42&52\\
&&5:23:43.7&-66:32:15.1&&20$\times$9.2&14&480&&&&&&\\
&&5:23:17&-66:27:45.3&&14$\times$9&11&650&&&&&&\\
&28&5:22:35&-66:28:52.8&&25$\times$17&21&1800&&17&63&0.25&60&73\\
&&5:23:17&-66:27:45.3&&14$\times$9&11&650&&&&&&\\
&29&5:25:33.3&-66:18:19.2&&34$\times$14&22&2600&&18&&&&100\\
&&5:25:9.7&-66:17:9.2&&40$\times$15&24&3300&&&&&&\\
&&5:24:52.9&-66:17:23.4&&8.6$\times$8.3&8.5&430&&&&&&\\
\hline
310.4&30&5:23:7.7&-66:32:12.9&&13$\times$8.5&10&350&&16&110&0.32&43&56\\
&&5:23:15.7&-66:28:31.1&&35$\times$20&26&2400&&&&&&\\
&&5:23:27.8&-66:25:7.4&&12$\times$7.5&9.6&450&&&&&&\\
&31&5:25:2.1&-66:22:58.7&&29$\times$19&23&2000&&22&140&0.61&71&89\\
&&5:24:12.7&-66:24:14.5&&37$\times$22&29&3100&&&&&&\\
&&5:23:26.3&-66:23:33.8&&16$\times$11&13&970&&&&&&\\
&32&5:25:12.5&-66:18:9.5&&21$\times$13&16&1600&&16&94&0.33&56&70\\
&&5:25:33.2&-66:17:30.1&&13$\times$11&12&840&&&&&&\\
&&5:26:11.1&-66:15:33.6&&31$\times$12&20&900&&&&&&\\

\hline
  \end{tabular}
  } 
 \end{center}
  \footnotetext{}{ (Continued Table)
	}
\end{table*}

\begin{table*}
 \caption{Filament Parameters}
 \label{c4_fil_para_4}
 \begin{center}
 \scalebox{0.8}{
  \begin{tabular}{lcllcccccccccc}
   \hline \hline

Channel &ID &\multicolumn{2}{c}{Position of the Cores}&&\multicolumn{3}{c}{Core parameters}&&\multicolumn{5}{c}{Filament Parameters}\\
 \cline{3-4} \cline{6-8} \cline{10-14}
Velocity&Number&R.A.($J$2000)&Dec($J$2000)&&axis $\sigma _{x}$$\times$$\sigma _{y}$&Diameter&Mass&&Width&Length&Mass&Line Mass 1&Line Mass 2\\
$[$km s$^{-1}]$&&h:m:s&d:$^{\prime}$:$^{\prime \prime}$&&[pc$\times$pc]&[pc]&[M$_{\odot}$]&&[pc]&[pc]&[10$^4$ M$_{\odot}$]&[M$_{\odot}$/pc]&[M$_{\odot}$/pc]\\
\hline 
&33&5:26:51.1&-66:10:45.4&&28$\times$21&24&1200&&21&120&0.32&38&53\\
&&5:27:10.6&-66:6:36.6&&30$\times$10&18&970&&&&&&\\
&&5:27:34.6&-66:4:9.8&&28$\times$15&20&1100&&&&&&\\
&34&5:26:7.5&-65:57:14.7&&70$\times$19&36&3500&&31&87&0.43&39&68\\
&&5:26:40&-65:52:13.4&&27$\times$13&19&740&&&&&&\\
&35&5:25:44.6&-66:11:46.8&&15$\times$7.6&11&620&&11&&&&58\\
&36&5:27:10.6&-66:0:49.7&&34$\times$13&21&480&&21&&&&23\\
&37&5:27:51&-66:7:25.7&&29$\times$8.1&15&590&&15&&&&38\\
\hline
315.4&38&5:23:11.8&-66:29:36.4&&33$\times$15&22&1700&&7.8&88&0.23&22&32\\
&&5:22:47.2&-66:29:29.3&&12$\times$5.9&8.6&140&&&&&&\\
&&5:22:34&-66:29:13.1&&13$\times$8.1&10&240&&&&&&\\
&&5:22:15.7&-66:27:59.4&&19$\times$7.7&12&170&&&&&&\\
&39&5:23:47.6&-66:27:47.2&&10$\times$4.9&7&170&&9.9&53&0.16&34&41\\
&&5:23:56.2&-66:26:19.3&&18$\times$9.2&13&410&&&&&&\\
&&5:24:9.3&-66:24:52.4&&16$\times$15&15&1000&&&&&&\\
\hline
  \end{tabular}
  } 
 \end{center}
  \footnotetext{}{ (Continued Table)
	}
\end{table*}

\subsection{Global Kinematics of the H$\,${\sc i} Ridge}

The global kinematics of the H$\,${\sc i} ridge provide clues to understanding the dynamics of its molecular cloud formation.
The three dimensional (position-position-velocity) distribution of the H$\,${\sc i} largely reflects the dominant dynamical processes occurring in the atomic ISM, such as gravitational infall, instability-induced collapse, and shock compression by the expanding shells.
We can therefore visualize these processes via position-velocity diagrams of the H$\,${\sc i} emission.

\subsubsection{Position-Velocity Diagrams: Perpendicular Cut}

\begin{figure*}
 \begin{center}
  \includegraphics[width=2\columnwidth]{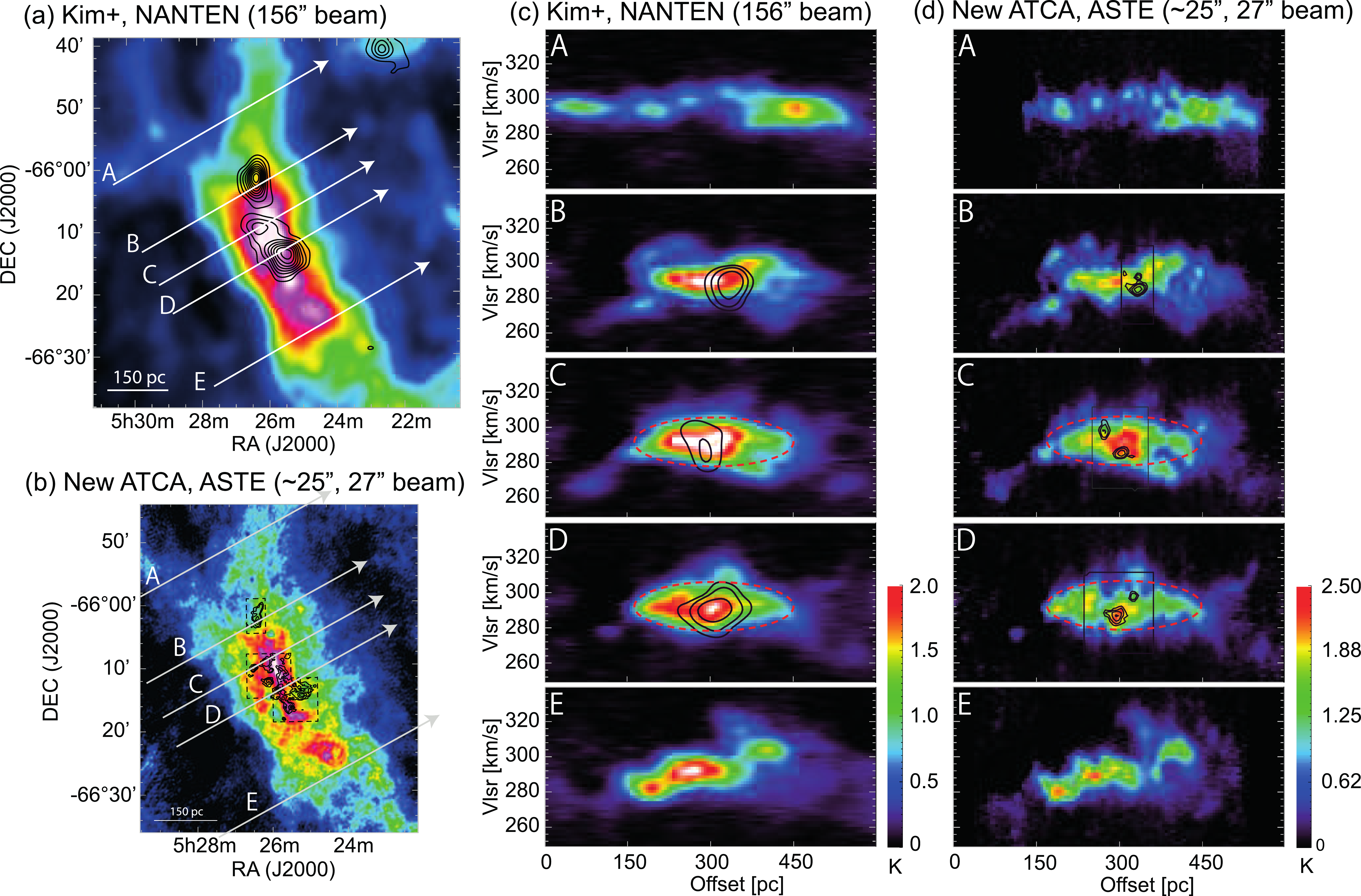}  
 \end{center}
\caption{
Position-velocity diagrams along cuts perpendicular to the ridge for the low resolution (156$^{\prime \prime}$) and new high resolution ($\sim 25 ^{\prime \prime}$) data. 
(a) Low-resolution map of the entire ridge showing the positions of the cuts (white arrows). Color is the H$\,${\sc i} of \citet{Kim_etal_2003} (smoothed to a resolution of 156$^{\prime \prime}$), and contours are NANTEN $^{12}$CO($J$=1--0) \citep{Fukui_etal_2008}. (b) The same map for the high resolution datasets. Color is the new H$\,${\sc i} data of the present work, and contours are ASTE $^{12}$CO($J$=3--2). 
(c) $P$--$V$ diagrams for the low resolution datasets. (d) $P$--$V$ diagrams for the high resolution datasets. Red dashed ellipses at positions C and D illustrate an axisymmetric, elliptical distribution in $P$--$V$ space. 
\label{fig_C4.HI.PV}}  

\end{figure*}

Figure \ref{fig_C4.HI.PV} shows position-velocity ($P$--$V$) diagrams of H$\,${\sc i} and CO along five cuts perpendicular to the direction of the ridge, allowing us to visualize its radial motion and the effects of the two SGSs.
The five cuts (labelled A to E) pass through typical regions at the northern and southern ends of the ridge (A and E), the peak position of the N49 GMC (B), and the two peak positions of the N48 GMC (C and D). 
Two extreme resolution images are presented ($\sim$ 156$^{\prime \prime}$ and $\sim$ 25$^{\prime \prime}$, corresponding to $\sim$ 38 pc and $\sim$ 6 pc), in order to probe both the global and local kinematics.  
For the low spatial resolution data, the archival H$\,${\sc i} data of \citet{Kim_etal_2003} and the NANTEN $^{12}$CO($J$=1--0) data of \citet{Fukui_etal_2008} are used,  
with the former smoothed 
to the NANTEN $^{12}$CO($J$=1--0) beam size ($\sim$ 156$^{\prime \prime}$). 
For the high spatial resolution map, our new ATCA H$\,${\sc i} data and the ASTE $^{12}$CO($J$=3--2) are used.
Since the difference in the beam size of these two datasets is small ($\sim$ 25$^{\prime \prime} \times 20 ^{\prime \prime}$ and $\sim$ 27$^{\prime \prime}$, $\sim$ 5--7 pc), no smoothing is performed.

In the low resolution $P$--$V$ diagrams, the H$\,${\sc i} is quite smoothly distributed, particularly at positions C and D, where the emission is concentrated in the central regions of the plot in an almost axisymmetric distribution, with no evidence of complex substructure. 
The peak velocity is roughly constant ($\sim$ 290 km s$^{-1}$) with direction -- i.e. there is no evidence of any shift in peak $V_{\rm lsr}$ from one side of the cut to the other.
The vast majority of the emission is contained in ellipses of $\sim$ 300 pc $\times$ 20 km s$^{-1}$ (as illustrated by red dashed lines in the figure).
As discussed in later, these features suggest that the external force of the expanding shells is not dominating the global kinematics of the ridge, and may provide tentative evidence that the system is now self-gravitating.
On the other hand, at positions A, B, and E, the H$\,${\sc i} shows several sub-components in $P$--$V$ space and the distribution is not ellipse-like.
Peak velocity varies with position, with more than two local peaks in positions A and E, and evidence of a systematic shift from $\sim$ 280 to 300 km s$^{-1}$ in positions B and E (moving from the LMC 4 to LMC 5 side of the ridge). 
The $^{12}$CO($J$=1--0) contours are roughly coincident with the H$\,${\sc i} peaks; on these size scales, CO and H$\,${\sc i} show good spatial agreement, and there is no evidence of H$\,${\sc i}-deficient positions that might indicate the conversion of H$\,${\sc i} to H$_2$.

In the high resolution $P$--$V$ diagrams complex sub-structure is revealed.
Even at positions C and D, the main H$\,${\sc i} component breaks down into several sub-components, which may correspond to structures such as the filamentary features discussed above -- also newly revealed at this improved spatial resolution.
Nevertheless, the overall distribution at these positions is still roughly centrally concentrated and axisymmetric (see again the red dashed ellipses in the figure).
The $^{12}$CO($J$=3--2) contours at positions B, C and D, resolve into a significantly more compact and clumped distribution. 
The molecular gas is still located within the atomic envelope, but is offset from the H$\,${\sc i} peaks in both the spatial and velocity directions (except for one clump at position D), 
demonstrating that dense molecular clumps need not always form at the center of the brightest atomic features.

\subsubsection{Position-Velocity Diagrams: Parallel Cut}
\begin{figure*}
 \begin{center}
  \includegraphics[width=2.1\columnwidth]{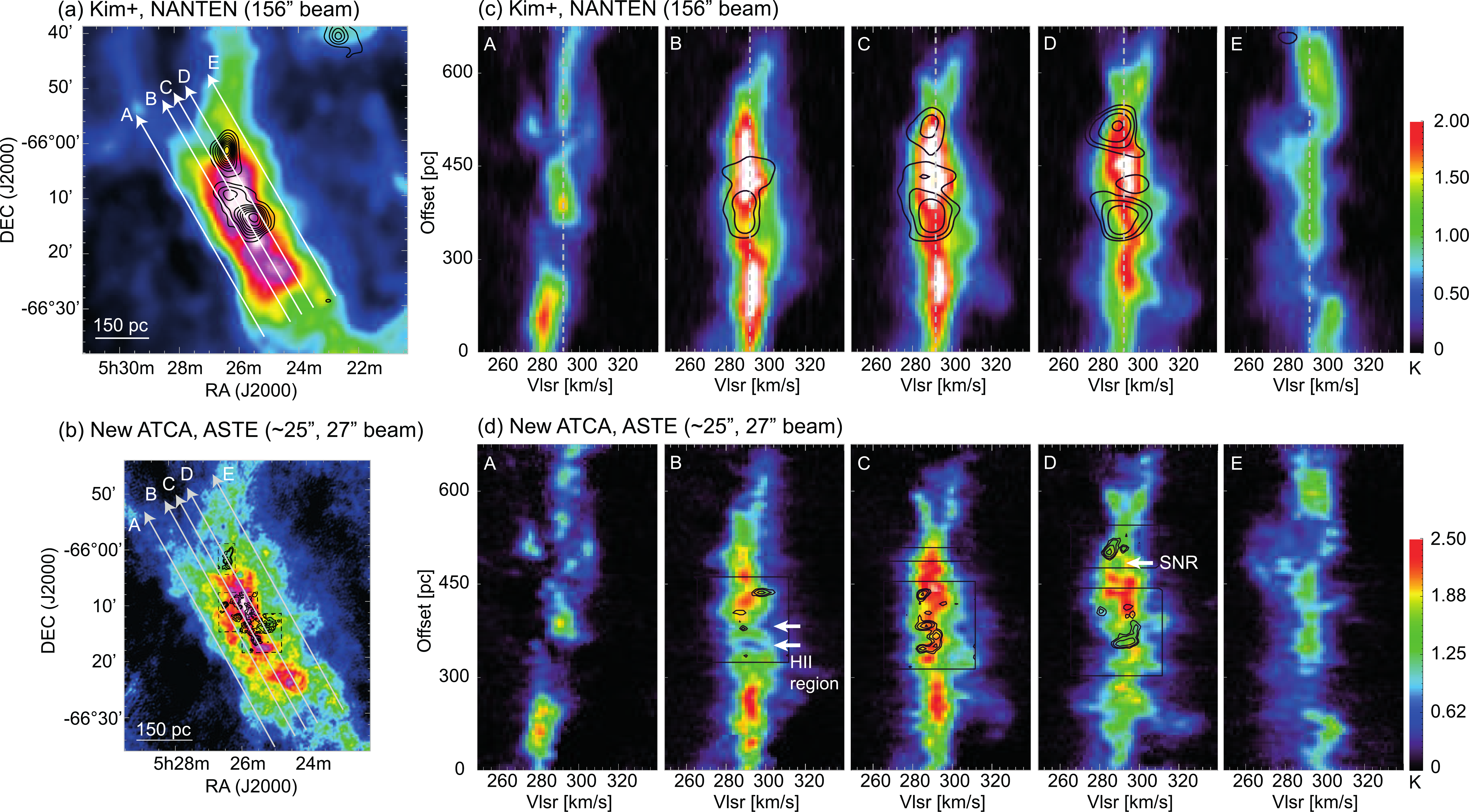}  
 \end{center}
\caption{
Position-velocity diagrams along cuts parallel to the ridge. All information is the same as in Figure \ref{fig_C4.HI.PV}. Grey dashed lines in the low resolution images mark $V _{\rm lsr}$ = 292 km s$^{-1}$.
\label{fig_C4.HI.PV.vDec}}  
\end{figure*}

Figure \ref{fig_C4.HI.PV.vDec} shows the $P$--$V$ diagrams in the direction parallel to the ridge, at the same resolutions and with the same datasets described above.
Cuts are performed at 5 positions (A to E in Figure \ref{fig_C4.HI.PV.vDec}).
Positions C and D pass through the strongest peaks of the N48 and N49 clumps, A and E pass through a diffuse part of the ridge, and B passes through the H$\,${\sc ii} region N48.

In the velocity cuts near the center of the ridge (B, C, and D), the low resolution H$\,${\sc i} is distributed in a straight line of roughly constant peak $V _{\rm lsr}$ spanning $\sim$ 600 pc in length.
To illustrate this, a dashed line of $V _{\rm lsr} = 292$ km s$^{-1}$ is drawn on the figure.
Although the peak $V _{\rm lsr}$ of the CO clouds are similar to the H$\,${\sc i}, the CO clouds are spatially offset from the strongest H$\,${\sc i} emission. 
The H$\,${\sc i} distribution in the diffuse portions of the ridge (A and E) appears somewhat more complex, with several clear sub-components, and a peak $V _{\rm lsr}$ that winds about the image.

The high resolution $P$--$V$ diagrams reveal considerably more complexity. 
At position C, the peak $V _{\rm lsr}$ shows several ``wiggles'', particularly around the N48 clumps at an offset position of $\sim$ 300 pc.
At position D, the molecular clumps are significantly offset from the atomic gas, with the CO distributed in several locations surrounding the brightest H$\,${\sc i}.
At position B, an H$\,${\sc i} intensity depression due to the H$\,${\sc ii} region N48 is clearly seen at offset position $\sim$ 250--300 pc. The depressed velocity range is quite wide (almost 20 km s$^{-1}$), indicating that the H$\,${\sc ii} region has effectively accelerated the gas surrounding it.
A neat H$\,${\sc i} hole is also seen around the SNR N49 at position D, at an offset position of $\sim$ 250--300 pc and $V _{\rm lsr} \sim$ 290 km s$^{-1}$.

\subsubsection{Interpretation of the $P$--$V$ Diagrams}
In the perpendicular cuts, the $P$--$V$ diagrams show a centrally-concentrated and axisymmetric H$\,${\sc i} distribution through the central portions of the ridge (positions C and D), but a more complex distribution in other areas (A, B, and E).
If the ridge is affected or pressure confined by the colliding shells, the distribution of emission in $P$--$V$ space is expected to show evidence of deformation on the edges bounding the shells, for example, several velocity components or a gradient in peak velocity. 
Similarly, if the ridge is composed of two shell walls immediately prior to collision, 
we might expect to see evidence of unique components associated with each shell at either end of the offset axis. 
These kinds of features are indeed seen at positions A, B, and E.
On the other hand, an isolated self-gravitating system should show a centrally-concentrated and axisymmetric distribution in $P$--$V$ space. This arises because gravity tends to act towards the center of a cloud, resulting in roughly axisymmetric kinematics with the highest column densities seen towards the cloud center.  
Any interaction between the two shells is expected to have proceeded furthest at the positions of the central cuts (C and D), where simple geometry suggests the shock fronts would have collided first. 
It is notable that these are the positions in which the $P$--$V$ diagrams show no significant sub-components, and at which no clear evidence of the effect of the shells remains (unlike positions A, B, and E).
Furthermore, the centrally-concentrated, axisymmetric distribution, with the GMCs located within the H$\,${\sc i} envelope, is close to that expected for a self-gravitating cloud.
We therefore suggest that the H$\,${\sc i} gas at positions C and D may now be well-mixed after the shell interaction, and is potentially now confined by self-gravity, while the ISM at positions A, B, and E may represent a less advanced stage of the shell interaction.


%% file: draft_05discussion.tex
High resolution ATCA$+$Parkes observations have revealed that the ISM ridge between the LMC supergiant shells LMC4 and LMC5 consists of a collection of H$\,${\sc i} filamentary features with characteristic widths of $\sim$20 pc.
In order to construct a suggested GMC formation scenario for this region, there are several questions to be addressed. 
\begin{itemize}
\item[\it Q1.] Can the accumulation of pure WNM by the expansion of the SGSs form the ridge, or is pre-existing dense material required?
\item[\it Q2.] Are the GMCs and the H$\,${\sc i} ridge gravitationally stable or not?
\item[\it Q3.] Does the characteristic separation of the clumps corresponds to the Jeans length?
\item[\it Q4.] How does the local structure of the filamentary features affect the GMC formation process?
\end{itemize}
In the following section, answers for these questions are outlined, and a formation scenario for the GMC and H$\,${\sc i} ridge are constructed.

\subsection{The Formation of the Ridge by the SGSs}

\cite{Dawson_etal_2015} have analyzed a GMC that is located at the interface of two Milky Way Supershells, GSH287$+$04$-$17 and the Carina OB2 Supershell.
These authors estimated a mean initial number density for the pre-shell medium $\left< n _{\rm H} \right>$, and concluded that the GMC was partially seeded by pre-existing material denser than the WNM, and assembled into its current form by the action of the two shells.
$\left< n _{\rm H} \right>$ was estimated by dividing the molecular mass of the GMC, $M _{{\rm H}_2}$, by the total volume of two cones with their apexes located at the shell centers, and their circular bases representing a disk-like GMC at the shell interface. 
In their case, the mass and the diameter of the cloud were $M _{{\rm H}_2} = 1.2 \times 10 ^5$ M$_{\odot}$ and 90 pc, and the heights of the cones ($\sim$  the radii of the shells) were 80 and 100 pc.
They found $\left< n _{\rm H} \right> \sim 10$ cm$^{-3}$, which is denser than a canonical ambient atomic medium ($n \sim$ 1 cm$^{-3}$), implying that some pre-existing dense material was present prior to the formation of the GMC -- either a mixture of WNM and CNM, or (more likely) also including some quantity of molecular gas. 

We now perform a similar estimation of the mean initial number density for the LMC H$\,${\sc i} ridge.
Note that in the present case the mass of the H$\,${\sc i} gas cannot be ignored. Taking the total H$\,${\sc i} mass of the ridge of 5$\times$10$^6$ M$_{\odot}$ (here the value is not opacity-corrected in order to compare with Kim et al.) into account, the total mass of the ridge $M$(H$\,${\sc i} + H$_2$) is $\sim$ $6.5 \times 10 ^6$ M$_{\odot}$. The base of the cones of swept-up material are estimated as ellipses with major and minor axes of 600 $\times$ 400 pc, which is roughly estimated elongation length and depth of the H$\,${\sc i} ridge (assuming the depth is similar to the H$\,${\sc i} scale hight of the LMC $\sim$ 180 pc; \citealt{Kim_etal_1999}).
With SGS radii of 700 pc (LMC 4) and 400 pc (LMC 5), $\left< n _{\rm H} \right> \sim 4$ cm$^{-3}$ is obtained. If the H{\sc i} opacity correction of Section \ref{s4_opacity} is used, then this mean number density increases to $\left< n _{\rm H} \right> \sim 6$ cm$^{-3}$

The average H$\,${\sc i} number density of the LMC is $\sim$ 2 cm$^{-3}$ (\citealt{Kim_etal_2003}). This can be assumed to consist of a mixture of WNM and CNM, where the contribution due to cold, optically thick H{\sc i} is somewhat underestimated due to the lack of opacity correction in the cited work. Thus our estimated initial density is comparable to that of ambient atomic medium of the LMC, and there is no need to invoke a significant quantity of pre-existing molecular gas. 
The large sizes of the SGSs means that they are able to accumulate material over very large distances, such that the large ridge mass of ($>$10$^6$ M$_{\odot}$) can in principle be 
accumulated from a purely atomic gas. Indeed, in order to form GMCs from purely atomic gas, such large-scale flows are required.

We note that numerical simulations of colliding flows in Milky Way-like conditions find that H$_2$ is formed in a few Myr, coagulating into larger H$_2$ clouds within $\sim$10 Myr \citep[e.g.][]{Clark_etal_2012,Inoue_inutsuka_2012}.
The typical lifetime of a super giant shell is $\sim$10--20 Myr (for the LMC see \citealt{Dopita_etal_1985, Points_etal_1999, Glatt_etal_2010}) as suggested by the stellar ages in the shells -- consistent with our picture of cloud formation within SGS lifetimes. 
The LMC's lower metallicity (0.3--0.5 Z$_{\odot}$; \citealt{Westerlund_1997}) and somewhat higher UV radiation field (\citealt{Israel_etal_1986}) may impact cloud formation timescales.
The lower metallicity in the LMC would prolong the H$_2$ formation timescale by roughly a factor of Z$_{\odot}$/Z$_{\rm{LMC}}$ (\citealt{Hollenbach_McKee_1979}), approximately doubling the formation time. In contrast, the impact of reduced UV shielding is expected to be less important, since self shielding is more important than dust shielding in H$_2$ (\citealt{Draine_Bertoldi_1996}).

\subsection{Gravitational Stability of the Clouds}
\label{grav_stab}
\cite{Dawson_etal_2015} also discussed the dominant confining pressure of the GMC located between GSH287$+$04$-$17 and the Carina OB2 Supershell.
The surface pressure just required to confine a uniform spherical cloud, with the inclusion of a surface pressure term in the virial theorem, can be obtained as
\begin{equation}
P _{S} = \frac{1}{4 \pi R ^3} \left( 3 M  \sigma _{v}^2 - \frac{3}{5} \frac{GM ^2}{R} \right).
\end{equation}
For the assumed properties of their GMC, \citet{Dawson_etal_2015} estimated that $P _{S} \gtrsim 7 \times 10^{-12}$ g cm$^{-1}$ s$^{-2}$ would be needed to confine the molecular gas in its current state, and suggested the ram pressure $\rho v ^2$ from the colliding flows as a candidate for this external confining pressure. From their accompanying hydrodynamic simulations, warm gas in the collision zone was typically found to have densities of 1--10 cm$^{-3}$ and velocities of 20--40 km s$^{-1}$ from each side, corresponding to ram pressures of $\sim 7 \times 10^{-12}$ and $3 \times 10^{-10}$ g cm$^{-1}$ s$^{-2}$ -- more than sufficient to confine the GMC. 

In the case of the GMCs in the LMC H$\,${\sc i} ridge, the situation is somewhat different.
Assuming a GMC radius of $R =50$ pc and taking the velocity width $dV =11.5$ km s$^{-1}$, the virial mass is $M _{\rm vir}$($^{12}$CO) $ \sim 1.3 \times 10^6$ M$_{\odot}$.
Given that the luminosity-based mass is $M _{\rm lum}$($^{12}$CO) $\sim 1.5 \times 10^6$ M$_{\odot}$, the virial parameter of the GMC in the ridge is $\alpha \sim 0.97$, indicating that self-gravity is sufficient to confine the molecular gas.
The surface pressure term $P _{S}$ is therefore negative for these GMC parameters, and no external pressure is necessary.

The gravitational confinement of the H$\,${\sc i} ridge also should be checked.
Assuming the main part of the H$\,${\sc i} ridge as a uniform sphere with 100 pc radius, and given the FWHM velocity width of 20 km s$^{-1}$, the virial mass of the ridge is $\sim$ 8$\times$10$^{6}$ M$_{\odot}$. Although the total mass of H$\,${\sc i} gas in the corresponding area is $\sim$ 3$\times$10$^{6}$ M$_{\odot}$, the total mass of the ridge including molecular mass (H$\,${\sc i}$+$H$_2$) is $\sim$ 5$\times$10$^{6}$ M$_{\odot}$ with an upper limit of $\sim$ 8$\times$10$^{6}$ M$_{\odot}$ (the opacity corrected value). This is slightly lower but comparable to the virial mass of the ridge. 
Considering that typical H$\,${\sc i} clouds are highly non-virialized (e.g., \citealt{Kim_etal_2007}), we can say that the H$\,${\sc i} ridge is almost virialized and is roughly confined by gravity as one self-gravitating system. 

Although the system as a whole is either roughly Virialized or slightly sub-Virialized, this does not imply that there is no infall of gas.
The computed Virial mass with the FWHM velocity width of the whole system reflects the global dynamics.
On local scales, individual structures may be more strongly gravitationally bound, and infall of atomic and/or molecular gas may be at work 
\citep[see e.g.][]{VzquezSemadeni_etal_2019}.
For example, sub-parsec resolution observations of the N159 region in the LMC with ALMA have revealed that young high-mass stars are formed with a mass accretion rate of $\sim$ 10$^{-4}$ M$_{\odot}$ yr$^{-1}$ (\citealt{Fukui_etal_2015a}), although the parental GMC as an entire system is roughly virialized.

We also note that the formation of self-gravitating molecular clouds via dynamical/thermal instabilities is consistent with simulations. 
Based on post-process Virial analysis, \citet{Inoue_inutsuka_2012} showed that molecular clouds created by a colliding flows can be observed as a virialized object even through their simulation does not model the self-gravity. The initial gas accumulation flow does not necessary need to be gravitationally driven in order to form self-gravitating clouds.

\subsection{Jeans Analysis \label{discussion_jeans}}

If the GMCs were formed via gravitational instability of the accumulated gas of the ridge, then the characteristic $\sim$ 40 pc separation of the molecular clumps should correspond to the Jeans length,
\begin{equation}
\lambda _{\rm J} = \sqrt{ \frac{\pi c _{\rm S} ^2 }{G \rho} },
\label{c5_thermjeans}
\end{equation}
\noindent which describes the critical size scale of a gravitationally unstable gas cloud.
Assuming that the kinetic and spin temperatures are equivalent, the sound speed $c _{\rm S}$ is then given by $c _{\rm S} = \sqrt{ \kappa R T _{\rm spin} / \mu }$, 
where $\kappa$ is the ratio of the specific heat ($\kappa = 5/3$ for the atomic medium) and $R$ is the gas constant.
Then Jeans mass $M _{\rm J}$ is then derived from
\begin{equation}
M _{\rm J} = \frac{4}{3} \pi \left( \frac{\lambda _{\rm J} } {2}  \right)^3 \mu \, m _{\rm H\,{\scriptstyle I}} \, n _{\rm H\,{\scriptstyle I}},
\end{equation}
where $\mu$ is the mean molecular weight (for the atomic medium $\mu \sim$1.3 including the mass of Helium), and $m _{\rm HI}$ is the mass of the hydrogen atom. 

Taking the initial condition of H$\,${\sc i} from the observed results above (see \S \ref{s4_opacity}), H$\,${\sc i} number density $n _{\rm HI} =$ 10 cm$^{-3}$ and temperature $T _{\rm spin} =$ 100--200 K, we obtain $\lambda _{\rm J} =$ 48--71 pc and $M _{\rm J} =$ 1.9--6.1 $\times 10^4$ M$_{\odot}$. 
The obtained $\lambda _{\rm J}$ and $M _{\rm J}$ for $T _{\rm spin} =$ 100 K (48 pc and 1.9 $\times 10^4$ M$_{\odot}$) are comparable to the observed clump separation $\sim$ 40 pc and the clump mass $\sim$ 2 $\times 10^4$ M$_{\odot}$.
However, $\lambda _{\rm J}$ and $M _{\rm J}$ for $T _{\rm spin} =$ 200 K are higher than the observed values.
Since $\lambda _{\rm J} \propto \sqrt{ T _{\rm spin} / n _{\rm HI}} $ and $M _{\rm J} \propto \sqrt{ T _{\rm spin}^{3} / n _{\rm HI} }$, a higher initial density would make $\lambda _{\rm J}$ and $M _{\rm J}$ more comparable to the observed separation and mass. For example, $n _{\rm HI} =$ 30 cm$^{-3}$ and $T _{\rm spin} =$ 100--200 K give $\lambda _{\rm J} =$ 27--41 pc and $M _{\rm J} =$ 1.0--3.5 $\times 10^4$ M$_{\odot}$, which are more comparable with the observed values. 
Taken together, these simple estimates imply that the N48 and N49 molecular clumps could have formed via gravitational instability in overdense and/or cold part of H$\,${\sc i} gas of the H$\,${\sc i} ridge, for example in the shocked H$\,${\sc i} gas most strongly impacted by the shells.

\subsection{Fragmentation of the Filamentary Features}

The structure of the H$\,${\sc i} ridge is dominated by filamentary features with characteristic widths of $\sim$ 20 pc (\S \ref{s4_filpara}).
The observed molecular clumps are located along prominent examples of these filamentary features, suggesting that the filamentary nature of the H$\,${\sc i} gas may play an important role in the molecular clump formation process. 

The characteristic isothermal scale-height $H$ of a gas cylinder in hydrostatic equilibrium is given by
\begin{equation}
H =\frac{ c_{\rm S} }{\sqrt{ 4 \pi G \rho }},
\label{c5_eq_filscale}
\end{equation}
where $c_{\rm S}$ is the sound speed, $G$ is the gravitational constant, and $\rho$ is the gas mass density at the center of the filament (e.g., \citealt{Nagasawa_1987}).
In this expression, the scale-height $H$ is equal to $\lambda _{\rm J, therm}/2 \pi$.
The width of the cylinder is $2H = \lambda _{\rm J, therm}/\pi$.
The characteristic fragmentation scale of a self gravitating cylinder is expected with separation of 4 times of the cylinder diameter (\citealt{Inutsuka_Miyama_1992}); i.e. the 
fragmentation wavelength is 
\begin{equation}
\lambda _{\rm frag} = 8H.
\label{c5_eq_filseparation}
\end{equation}

The cases with density of 10 cm$^{-3}$ and temperatures of 100 to 200 K give widths of $\sim$ 15--22 pc ($=2$H), which is in good agreement with the observed widths of the filamentary features. This suggests that the filamentary features may be in a state close to hydrostatic equilibrium.
The implications of the Jeans analysis above also apply in this case: when the H$\,${\sc i} gas is denser or cooler, $\lambda _{\rm frag}$ becomes smaller than the observed width. Since most of the filamentary features are not associated with molecular clumps, the physical parameters of the H$\,${\sc i} filamentary features must not in general be sufficient for molecular clump formation. This implies that the filamentary features need to be denser to form molecular clumps, and again, that the N48 and N49 molecular clumps may have form in shocked parts of H$\,${\sc i} gas, most strongly impacted by the action of the shells.

\subsection{The GMC Formation Scenario at the Ridge}
The answers to the questions posed at the start of this section to explore the GMC formation scenario in the H$\,${\sc i} ridge are as follows.
\begin{itemize}
\item[\it A1.] The total mass of the ridge can be supplied by the accumulation of the ambient atomic medium by the two SGSs. The required initial number density is a few cm$^{-3}$, and as such does not require the existence of any significant quantities of pre-existing molecular gas.
In order to form GMCs of $\sim$10$^6$ M$_{\odot}$ from a purely atomic medium, indeed such kilo-parsec scale flows are required.
\item[\it A2.] Both the GMCs and the H$\,${\sc i} ridge are roughly in virial equilibrium. We suggest that the entire system, including the sub-structures of the clumps, may be well explained as a single self-gravitating system. 
\item[\it A3.] 
The separation and the mass of the N48 clumps can be explained by Jeans length and Jeans mass. 
This follows a scenario in which the molecular clumps formed via gravitational instability in the material accumulated by the two shells.  
\item[\it A4.] The molecular clumps are located along prominent filamentary features, which suggests that the filamentary structure of the atomic gas may play an important role in the formation of the molecular clumps. 
\end{itemize}

Considering these facts, a scenario for the GMC formation process in the ridge can be constructed. 
On large ($>$100 pc) scales, the expansion and subsequent collision of the shells accumulates the ambient atomic medium, with an initial density of a few cm$^{-3}$, into a large, high column density ridge that mainly consists of shocked CNM with a density of several tens of cm$^{-3}$. 
The ridge then collapses under gravity into GMCs consisting of many molecular clumps. These clumps have typical masses of $\sim$ 10$^4$ M$_{\odot}$ and densities of $\sim$ 10$^3$ cm$^{-3}$.
At high-resolutions ($<$10 pc), 
the structure of the H$\,${\sc i} gas becomes filamentary during its evolution, and the formation of the molecular clumps occurs along atomic filaments. 
These formation scenarios proceed simultaneously at both size scales. 
This picture agrees well with theoretical predictions suggesting that several episodes of compression are required to form GMCs, and newly suggests that GMC formation involves the filamentary nature of the atomic medium.

\subsection{Speculation on the General GMC Formation Process}

Since the distribution of filamentary H$\,${\sc i} features is quite crowded in the ridge, the merging of the filamentary features induced by the shell collision might play an important role in increasing their mass and density. In the LMC N159W and N159E regions, sub-parsec resolution observations with ALMA have revealed that high mass young stars are formed at the intersection of filamentary molecular clouds (\citealt{Fukui_etal_2015a, Saigo_etal_2016}). If the molecular clumps are formed in merging filaments, this would present a hierarchical view of the evolution of the ISM; the collision of filamentary H$\,${\sc i} clouds forms molecular clouds, and the collision of filamentary molecular clouds forms massive stars. To probe such a scenario, a large sample of detailed H$\,${\sc i} observations are required. It is also important to investigate the molecular clumps of the N48 and N49 regions in more detail, in order to understand whether the filamentary H$\,${\sc i} gas may be the origin of filamentary molecular clouds, like those seen in N159W.

%% file: draft_06summary.tex
The high column density H$\,${\sc i} ridge between the two kpc-scale supergiant shells, LMC4 and LMC 5, has been analyzed via new high-resolution observations 
made with the ATCA. The GMC formation process in the collision zone has been studied at multiple size-scales, from fine-structure ($<$ 10 pc) to the large-scale kinematics of the H$\,${\sc i} gas.
The main results and suggestions are as follows:
\begin{enumerate}
\item By combining new, ATCA long-baseline H$\,${\sc i} 21 cm line observations with archival, short-baseline data from the same instrument, and single dish data from the Parkes 64m Dish, a synthesized beam size of 24.75$^{\prime \prime}$ by 20.48$^{\prime \prime}$ ($\sim$ $6 \times 5$ pc at the LMC) was achieved. This is an unusually high spatial resolution for the 21 cm line in external galaxies. These new observations reveal that the structure of the H$\,${\sc i} gas is highly filamentary.
\item The H$\,${\sc i} opacity correction method of \cite{Fukui_etal_2014b, Fukui_etal_2015a} was applied to the H$\,${\sc i} ridge, in the first test of this method in an external galaxy. 
After opacity correction, a rough upper limit on the total mass of the H$\,${\sc i} ridge is estimated as $\sim 8.5 \times 10^6$ M$_{\odot}$, which corresponds to a factor of 1.7 times greater than the uncorrected data. The harmonic mean optical depth and spin temperature are estimated to be $\tau _{\rm HI} \gtrsim 1.5$ and $T _{S} \gtrsim 100$ K.
\item We identify filamentary features in the atomic medium by connecting H$\,${\sc i} cores.
In total 39 features are identified, implying that the H$\,${\sc i} gas structure of the ridge mainly consists of a collection of filamentary features.
The features have typical widths of $\sim$  21 (8--49) [pc], and line masses of $\sim$  90 (20--190) [M$_{\odot}$/pc].
\item H$\,${\sc i} position-velocity diagrams perpendicular to the ridge show a centrally-concentrated and axisymmetric distribution in the shell collision zone (the N48 region), with the molecular clouds located close to the central regions.
The H$\,${\sc i} gas may now be well-mixed after the shell interaction.
\item A scenario for the GMC formation process in the N48 and N49 regions can therefore be constructed as follows:
On large-scales, the expansion of the two SGSs accumulates the atomic medium into a high column density ridge. 
The ridge collapses into GMCs consisting of molecular clumps.
On $\sim$10 pc scales, the H$\,${\sc i} becomes highly filamentary during its evolution. The molecular clumps are formed along the evolved filamentary H$\,${\sc i} gas. 
This picture agrees well with theoretical predictions that several episodes of compression are required to form GMCs, and also suggest that GMC formation is related to the filamentary nature of the atomic medium.
\end{enumerate}

\section{Acknowledgments}
A part of this study was financially supported by MEXT Grants-in-Aid for Scientific Research (KAKENHI) Grant Numbers 15071202, 15071203, 20001003, and JSPS KAKENHI Grant Numbers 14J11419, 22740127.
JRD is the recipient of an Australian Research Council DECRA Fellowship (project number DE170101086).
The Australia Telescope Compact Array is part of the Australia Telescope National Facility which is funded by the Australian Government for operation as a National Facility managed by CSIRO.
The ASTE project is managed by Nobeyama Radio Observatory (NRO), a branch of the National Astronomical Observatory of Japan (NAOJ), in collaboration with University of Chile, and Japanese institutes including University of Tokyo, Nagoya University, Osaka Prefecture University, Ibaraki University, and Hokkaido University. Observations with ASTE were carried out remotely from Japan by using NTT's GEMnet2 and its partner R\&E (Research \& Education) networks, which are based on the AccessNova collaboration of University of Chile, NTT Laboratories, and NAOJ.
The Mopra Telescope is part of the Australia Telescope which is funded by the Commonwealth of Australia for operation as a National Facility managed by CSIRO.
SAGE research has been funded by NASA/Spitzer grant 1275598 and NASA NAG5-12595.
Cerro Tololo Inter-American Observatory (CTIO) is operated by the Association of Universities for Research in Astronomy Inc. (AURA), under a cooperative agreement with the National Science Foundation (NSF) as part of the National Optical Astronomy Observatories (NOAO). The MCELS is funded through the support of the Dean B. McLaughlin fund at the University of Michigan and through NSF grant 9540747.

\section{Data availability}
The data underlying this article will be shared on reasonable request to the corresponding author.